\documentclass[
aps,
pra,
reprint,
superscriptaddress,
amsmath,
amssymb,
nofootinbib,
floatfix
]{revtex4-2}

\usepackage{graphicx}
\usepackage{bm}
\usepackage{dcolumn}
\usepackage{tikz}
\usepackage{etoolbox}
\usepackage{xurl}

\makeatletter
\patchcmd{\@bibdataout@aps}{author="08"}{author="48"}{}{}
\makeatother

\newcommand{\figpanel}[3][0.13,0.97]{%
\begin{tikzpicture}
\node[anchor=south west,inner sep=0] (img) at (0,0) {\includegraphics[width=0.49\textwidth]{#2}};
\begin{scope}[x={(img.south east)},y={(img.north west)}]
\node[anchor=north west,font=\footnotesize] at (#1) {(#3)};
\end{scope}
\end{tikzpicture}}

\begin{document}

\title{Self-Energy Construction for Chain Representations of Open Quantum Systems}

\author{Pengju Chen}
\email{pchen9@stevens.edu}
\affiliation{Center for Quantum Science and Engineering, Department of Physics, Stevens Institute of Technology, Hoboken, New Jersey 07030, USA}

\author{Da-Wei Luo}
\affiliation{Center for Quantum Science and Engineering, Department of Physics, Stevens Institute of Technology, Hoboken, New Jersey 07030, USA}

\author{Ting Yu}
\affiliation{Center for Quantum Science and Engineering, Department of Physics, Stevens Institute of Technology, Hoboken, New Jersey 07030, USA}

\date{\today}

\begin{abstract}
We identify the self-energy as the environmental response to the system, which transforms covariantly under shifts of the system--environment cut.
For a semi-infinite chain with asymptotically uniform parameters, this transformation admits a fixed point that defines the asymptotic boundary self-energy (BSE), whose spectral density is the Wigner semicircle.
A chain-extended BSE exploits the trend of the retained chain, offering improved accuracy in representing the environment.
The transformation introduces a pseudo-hyperbolic self-energy distance that is invariant under shifts of the cut, providing a natural geometry for comparing environmental responses.
As a demonstration, this geometry enables cut-independent error measure of finite damped-oscillator closures of the chain, distinguishing the closure error from the chain depth error, and offers a more accurate fitting scheme than fitting the time-domain correlation function.
\end{abstract}

\maketitle

\section{Introduction}

In open quantum systems, accurate and efficient representation of the environment is key.
In our view, this is related to the partitioning of the total system into a system of interest and an environment~\cite{breuer2002theory,gardiner2004quantum}, which is not unique.
Open-system dynamics can be treated using projection-operator methods~\cite{nakajima1958quantum,zwanzig1960ensemble} and influence-functional approaches~\cite{FEYNMAN1963118}, while quantum dynamical semigroups provide a framework for Markovian dynamics~\cite{gorini1976completely,lindblad1976generators}.
An environment is commonly modeled as a bath of harmonic oscillators linearly coupled to the system, with its mode frequencies and coupling strengths characterized by a spectral density~\cite{breuer2002theory,sanchez2021accurate}.
Structured spectral densities can give rise to non-Markovian dynamics~\cite{rivas2014quantum,breuer2016colloquium,de2017dynamics}, which can assist entanglement generation~\cite{zhao2011dynamics} and preservation~\cite{mu2016memory}, modify geometric phases~\cite{luo2018memory}, and support coherence~\cite{xiong2015non}.
Methods for studying non-Markovian effects include non-Markovian quantum-state diffusion~\cite{yu1999non,strunz2004convolutionless,chen2014exact,xu2014perturbation}, as well as exact master-equation and Green-function approaches~\cite{hu1992quantum,an2007non,zhang2012general,zhang2019exact,xu2026colloquium}.
Structured environments can produce non-Markovian dynamics when the system couples indirectly to a Markovian reservoir~\cite{ma2014crossover}.
This illustrates the non-uniqueness of the system--environment partition, where the effective environmental response depends on the choice of the system--environment cut~\cite{diosi1998non}.

Reaction-coordinate mappings reorganize the partition~\cite{garg1985effect,iles2014environmental,strasberg2016nonequilibrium,nazir2019reaction}.
Chain mappings allow the partition to be placed after any chain site~\cite{chin2010exact,martinazzo2011communication,woods2014mappings,huh2014linear}, and the chain beyond the partition is described by a residual spectral density.
For a single spectral band satisfying the convergence conditions of Ref.~\cite{woods2014mappings}, this residual spectral density approaches the Wigner semicircle in the asymptotic region~\cite{woods2014mappings,nusseler2022fingerprint}.
This chain representation supports efficient simulations based on time-dependent density-matrix renormalization group methods~\cite{prior2010efficient,de2015thermofield,tamascelli2019efficient}.
In these simulations, the chain beyond the partition can be replaced by a finite set of damped modes~\cite{sanchez2021accurate,nusseler2022fingerprint}.
We describe the chain beyond the partition by the environmental self-energy, which is related to the spectral density through a Cauchy transform.

Prior work has established the mathematical basis for the self-energy description of the environmental response.
Harmonic systems linearly coupled to harmonic environments admit exact descriptions for general spectral densities and temperatures~\cite{hu1992quantum,halliwell1996alternative,chou2008quantum,chou2008exact,martinez2013dynamics}.
Environmental self-energies have been developed in Green-function formulations~\cite{jauho1994time,dhar2006heat,jin2010non,zhang2012general}, which have also been used to analyze crosstalk in superconducting qubits~\cite{alghadeer2026crosstalk}.
Continued-fraction representations of environmental responses have been established for harmonic environments~\cite{hughes2009effective,martinazzo2011communication,martinazzo2011unraveling,woods2014mappings} and quantum circuits~\cite{bakr2026exact}.
Each step of the continued-fraction recurrence corresponds to a M\"obius transformation of the self-energy.
In quantum transport, the self-energy that replaces a semi-infinite contact in the partitioning approach~\cite{caroli1971direct} has been called the boundary self-energy~\cite{lake1997single,luisier2006atomistic,luisier2009atomistic}.

In this paper, we identify the self-energy as the environmental response and show that its continued-fraction recurrence defines a covariant transformation under shifts of the system--environment cut.
When this transformation law is applied to the asymptotic region of a semi-infinite lossless chain representation of the environment~\cite{woods2014mappings}, it provides an analytic boundary self-energy (BSE), defined as the self-energy whose recurrence closes on itself.
The BSE has the form of the square-root terminator of the recursion method~\cite{haydock1972electronic,haydock1980recursive,beer1984recursion}, and its real part on the real-frequency axis follows the Wigner semicircle law~\cite{woods2014mappings,nusseler2022fingerprint}.
We extend the idea of the BSE to match the trend of the retained chain, obtaining a chain-extended BSE that offers higher accuracy than the asymptotic BSE.
The same transformation introduces a pseudo-hyperbolic geometry that gives a cut-invariant distance between two environmental responses.
This can provide a basis for a geometric approach to understanding and engineering structured environments.
As a demonstration, this distance is used as an error measure for fitting the coefficients of the Markovian closure (MC), a finite damped harmonic oscillator closure of the chain representation~\cite{nusseler2022fingerprint}.
Our results show that this measure improves on the previous time-domain fitting strategy, and that the chain-extended BSE also offers a better target than the asymptotic BSE.

This paper is organized as follows.
Section~\ref{System and boundary self-energy} introduces the qubit--chain model and derives the self-energy transformations, the asymptotic BSE, and the chain-extended BSE.
Section~\ref{sec:termination_methods} relates the self-energy to the spectral density, describes the semicircular tail and the MC, and develops the cut-invariant distance.
Section~\ref{Numerical_demonstration} numerically compares different closures of the semi-infinite chain.
Sections~\ref{Discusses} and~\ref{Conclusion} present the discussion and conclusion, respectively.
Appendix~\ref{app:bse_recurrence} gives the Green-function representation and the derivation of the self-energy recurrences, while Appendix~\ref{app:star_chain_bse} shows the star--chain equivalence in terms of self-energy and the construction of the chain coefficients.

\section{System and self-energy}\label{System and boundary self-energy}
\begin{figure}[htbp]
\centering
\includegraphics[width=\columnwidth]{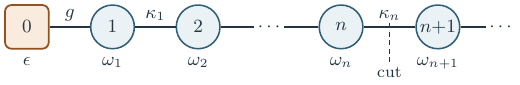}
\caption{Qubit (site $0$) coupled to a semi-infinite chain of harmonic oscillators.
The site frequencies are shown below the nodes and the couplings above the links.
The dashed line marks a possible system--environment cut between sites $n$ and $n+1$.}
\label{fig:chain_topology}
\end{figure}

We begin with a qubit coupled to a semi-infinite chain of harmonic oscillators.
The Hamiltonian is ($\hbar=1$)
\begin{equation}\label{eq:H_F}
\begin{split}
H&=\epsilon\sigma_+\sigma_-+g(\sigma_+a_1+\sigma_-a_1^\dagger)
\\
&+\sum_{n=1}^{\infty}\omega_n a_n^\dagger a_n +\sum_{n=1}^{\infty}\kappa_n\left(a_n^\dagger a_{n+1}+a_{n+1}^\dagger a_n\right).
\end{split}
\end{equation}
Here, $\epsilon$ is the qubit transition frequency, and we label the qubit as site $0$ and the harmonic oscillators as sites $n\geq1$.
The coupling $g$ describes excitation exchange between the qubit and the first oscillator.
The operator $a_n^\dagger$ ($a_n$) creates (annihilates) an excitation in the bath mode at site $n$, while $\omega_n$ is the corresponding mode frequency and $\kappa_n$ is the nearest-neighbor coupling between sites $n$ and $n+1$.
For environments with finite spectral support $[\omega_L,\omega_R]$ satisfying the convergence conditions of Ref.~\cite{woods2014mappings}, $\omega_n$ and $\kappa_n$ approach stable asymptotic values as $n\to\infty$.
Figure~\ref{fig:chain_topology} shows the chain and a possible system--environment cut.

To illustrate the dynamics, we consider the vacuum and single-excitation sectors, in which the total state can be written as
\begin{equation}\label{eq:total_state}
|\psi(t)\rangle = A_0(t)|1_q,\mathbf 0\rangle +B(t)|\text{vac}\rangle +\sum_{n=1}^{\infty}A_n(t)|0_q,1_n\rangle,
\end{equation}
where $|\mathbf 0\rangle$ denotes the chain vacuum, $|1_q\rangle$ and $|0_q\rangle$ denote the excited and ground states of the qubit, respectively, and $|1_n\rangle$ denotes a single excitation at chain site $n$.
The total vacuum is $|\text{vac}\rangle\equiv|0_q,\mathbf 0\rangle$.
In the following, we denote the single-excitation basis states by $|n\rangle$.
In particular, $|0\rangle\equiv|1_q,\mathbf 0\rangle$ represents the excited qubit with the chain in vacuum, while for $n\ge1$, $|n\rangle\equiv|0_q,1_n\rangle$ represents the qubit in its ground state with one excitation at chain site $n$.
Here, $A_0(t)$ and $A_n(t)$ are the corresponding excitation amplitudes, and $B(t)$ is the global vacuum amplitude.
We assume that the qubit and the chain are initially uncorrelated.

The vacuum amplitude $B(t)$ is decoupled from the dynamics of $A_0(t)$ and $A_n(t)$ (see Appendix~\ref{app:bse_recurrence}).
We write the excitation amplitudes as
\begin{equation}
\mathbf A(t) =\left(A_0(t), A_1(t),  A_2(t),  \cdots\right)^{T}.
\end{equation}
In this basis, the Schr\"odinger equation leads to
\begin{equation}
i\frac{d}{dt}\mathbf A(t) = \mathcal J_{\mathrm H}\mathbf A(t),
\end{equation}
where $\mathcal J_{\mathrm H}$ is the Jacobi matrix representation of the Hamiltonian in Eq.~\eqref{eq:H_F} restricted to the single-excitation sector,
\begin{equation}
\mathcal J_{\mathrm H} =
\begin{pmatrix}
\epsilon & g & 0 & 0 & \cdots \\
g & \omega_1 & \kappa_1 & 0 & \cdots \\
0 & \kappa_1 & \omega_2 & \kappa_2 & \cdots \\
0 & 0 & \kappa_2 & \omega_3 & \cdots \\
\vdots & \vdots & \vdots & \vdots & \ddots
\end{pmatrix}.
\end{equation}
Then, defining the Laplace transform $\boldsymbol{\mathcal A}(p)=\mathcal L[\mathbf A(t)]$, the Laplace-domain amplitudes satisfy
\begin{equation}
\boldsymbol{\mathcal A}(p)=\left(pI+i\mathcal J_{\mathrm H}\right)^{-1} \mathbf A(0),
\end{equation}
with $\operatorname{Re}p>0$.
For a chain initially in vacuum and a qubit state $c_e|1_q\rangle+c_g|0_q\rangle$, with $|c_e|^2+|c_g|^2=1$, the initial amplitudes are
\begin{equation}
\mathbf A(0)=\left(c_e,0,0,\cdots\right)^T, \qquad B(0)=c_g.
\end{equation}
For simplicity, we take the qubit initially excited, corresponding to $c_e=1$ and $c_g=0$.

In this one-dimensional chain, an oscillator site $n$ has two boundary ports, connecting it to its nearest-neighbor sites $n-1$ and $n+1$.
We label the corresponding ports by $n-1\to n$ and $n+1\to n$, where the arrow points from the side whose response is presented toward site $n$.
For the $n-1\to n$ port with $n\geq 2$, the \textbf{self-energy} is defined as the response of all degrees of freedom on the $n-1$ side of the cut as seen from site $n$,
\begin{equation}\label{eq:main Schur BSE left}
\begin{split}
&\Sigma_{n-1\to n}(p)
\\
&=\kappa_{n-1}^2\left\langle n-1\left| \left(pI_{0:n-1}+i\mathcal{J}_{0:n-1}\right)^{-1} \right|n-1\right\rangle,
\end{split}
\end{equation}
where $\mathcal{J}_{0:n-1}$ is the Jacobi matrix of the structure extending from the qubit through site $n-1$ in the single-excitation sector, and the identity matrix $I_{0:n-1}$ matches its dimension.

The Schur-complement reduction gives the recurrence relation
\begin{equation}\label{eq:main BSE recurrence left}
\Sigma_{n-1\to n}(p)=\frac{\kappa_{n-1}^2}{p+i\omega_{n-1}+\Sigma_{n-2\to n-1}(p)}.
\end{equation}
At the qubit endpoint, the recurrence starts from
\begin{equation}
\Sigma_{0\to1}(p)=\frac{g^2}{p+i\epsilon}.
\end{equation}
For $\kappa_{n-1}\neq0$, Eq.~\eqref{eq:main BSE recurrence left} is an invertible M\"obius transformation of the scalar self-energy.
Its inverse is
\begin{equation}\label{eq:inverse BSE recurrence left}
\Sigma_{n-2\to n-1}(p)=\frac{\kappa_{n-1}^2}{\Sigma_{n-1\to n}(p)}-p-i\omega_{n-1}.
\end{equation}
Similarly, for the $n+1\to n$ port, the self-energy is
\begin{equation}\label{eq:main Schur BSE right}
\begin{split}
&\Sigma_{n+1\to n}(p)
\\
&=\kappa_n^2 \left\langle n+1 \left| \left(pI_{n+1:\infty}+i\mathcal{J}_{n+1:\infty}\right)^{-1} \right|n+1 \right\rangle,
\end{split}
\end{equation}
where $\mathcal{J}_{n+1:\infty}$ is the Jacobi matrix of the semi-infinite structure beginning at site $n+1$.
The corresponding recurrence relation is
\begin{equation}\label{eq:main BSE recurrence right}
\Sigma_{n+1\to n}(p)=\frac{\kappa_n^2}{p+i\omega_{n+1}+\Sigma_{n+2\to n+1}(p)}.
\end{equation}
For $\kappa_n\neq0$, the inverse transformation is
\begin{equation}\label{eq:inverse BSE recurrence right}
\Sigma_{n+2\to n+1}(p)=\frac{\kappa_n^2}{\Sigma_{n+1\to n}(p)}-p-i\omega_{n+1}.
\end{equation}
The self-energies in Eqs.~\eqref{eq:main Schur BSE left} and~\eqref{eq:main Schur BSE right} are the corresponding surface Green functions of the structures on either side of the cut, multiplied by the squared coupling coefficients~\cite{lake1997single,luisier2006atomistic}.
Equation~\eqref{eq:main BSE recurrence right} gives the continued-fraction representation of the environmental response~\cite{hughes2009effective,martinazzo2011communication,martinazzo2011unraveling,woods2014mappings}.
The right-tail self-energy $\Sigma_{n+1\to n}$ depends only on the harmonic bath and its coupling to the retained site, so its recurrence does not require the single-excitation restriction.
The recurrences in Eqs.~\eqref{eq:main BSE recurrence left} and~\eqref{eq:main BSE recurrence right} are covariant in the sense that the self-energy keeps the same recursive form under shifts of the system--environment cut.
The derivation of Eqs.~\eqref{eq:main Schur BSE left}--\eqref{eq:inverse BSE recurrence right} is given in Appendix~\ref{app:bse_recurrence}.

The two recurrence relations describe how the system--environment boundary moves when a site is eliminated.
For Eq.~\eqref{eq:main BSE recurrence left}, the side containing the qubit is treated as the environment and the semi-infinite chain side as the system.
Transforming from $\Sigma_{n-2\to n-1}(p)$ to $\Sigma_{n-1\to n}(p)$ means eliminating site $n-1$, which moves the system--environment boundary toward increasing site index and produces $\Sigma_{n-1\to n}(p)$ as the environmental response presented to site $n$.
For Eq.~\eqref{eq:main BSE recurrence right}, the qubit side is treated as the system and the semi-infinite chain side as the environment.
Transforming from $\Sigma_{n+2\to n+1}(p)$ to $\Sigma_{n+1\to n}(p)$ means eliminating site $n+1$, which moves the boundary toward decreasing site index and produces $\Sigma_{n+1\to n}(p)$ as the environmental response presented to site $n$.
Thus, the local Schur-complement transformations propagate the self-energy in either direction, depending on which side's degrees of freedom we choose to eliminate.

At a finite site index $n$, the two boundary ports generally see different self-energies, i.e.,
\begin{equation}
\Sigma_{n-1\to n}(p)\neq\Sigma_{n+1\to n}(p).
\end{equation}
In the asymptotic region, however, the chain becomes locally homogeneous, with
\begin{equation}
\omega_n\to\omega_\infty,\qquad\kappa_n\to\kappa_\infty.
\end{equation}
The self-energy limits below are taken at fixed $p$ with $\operatorname{Re}p>0$, for which the influence of the remote inhomogeneous structure becomes asymptotically small in the continued fraction.
The two boundary responses then approach the same asymptotic fixed point,
\begin{equation}
\lim_{n\to\infty}\Sigma_{n-1\to n}(p)=\lim_{n\to\infty}\Sigma_{n+1\to n}(p)=\Sigma_{\infty}(p).
\end{equation}
At this fixed point, moving the boundary by one additional site leaves the asymptotic self-energy unchanged, so the recurrence relation closes on itself,
\begin{equation}\label{eq:self referring}
\Sigma_{\infty}(p)=\frac{\kappa_\infty^2}{p+i\omega_\infty+\Sigma_{\infty}(p)}.
\end{equation}
Solving the fixed-point equation gives
\begin{equation}\label{BSE_closure}
\Sigma_\infty(p)=\frac{-(p+i\omega_\infty)+\sqrt{(p+i\omega_\infty)^2+4\kappa_\infty^2}}{2}.
\end{equation}
Since the $p$ term dominates each denominator for large $|p|$ in the continued fraction, the self-energy must vanish as $\Sigma_\infty(p)\sim\kappa_\infty^2/p$.
We choose the quadratic root in Eq.~\eqref{BSE_closure} that is analytic for $\operatorname{Re}p>0$ and satisfies this large-$|p|$ behavior.
We denote $\Sigma_\infty$ as the asymptotic boundary self-energy (BSE).
In the following, BSE refers to $\Sigma_\infty$ unless otherwise specified.
An equivalent construction is given in Ref.~\cite{woods2014mappings} through spectral measures, and the BSE defined in Eq.~\eqref{BSE_closure} reproduces the Wigner semicircle residual spectral density (see Sec.~\ref{sec:termination_methods}).

We then choose $N$ sufficiently large that the chain coefficients are close to their asymptotic values, approximate $\Sigma_{N+2\to N+1}(p)$ by $\Sigma_\infty(p)$, and eliminate site $N+1$ to obtain the self-energy presented to site $N$,
\begin{equation}\label{time-nonlocal_bse_p}
\Sigma_{N+1\to N}^{\rm BSE}(p)=\frac{\kappa_N^2}{p+i\omega_{N+1}+\Sigma_\infty(p)}.
\end{equation}
We call this way of terminating the semi-infinite chain the BSE closure.
The first $N$ sites encode the ``fingerprint'' of the environment~\cite{nusseler2022fingerprint}, while the BSE represents the universal closure defined by $\{\omega_\infty,\kappa_\infty\}$.

The BSE $\Sigma_\infty$ approximates the exact self-energy $\Sigma_{N+2\to N+1}$ in the asymptotic region, which leaves room for adjusting its parameters to improve the accuracy.
Consider a chain-extended BSE $\bar{\Sigma}_N(p)$, which keeps the form of $\Sigma_\infty(p)$ with parameters determined from the available chain coefficients,
\begin{equation}\label{Extended_BSE_closure}
\bar{\Sigma}_N(p)=\frac{-(p+i\bar{\omega}_N)+\sqrt{(p+i\bar{\omega}_N)^2+4\bar{\kappa}_N^2}}{2},
\end{equation}
where $\bar{\omega}_N$ and $\bar{\kappa}_N$ are the effective tail frequency and nearest-neighbor coupling, given by
\begin{equation}
\bar{\omega}_N=\omega_\infty+\frac{a}{(N+1)^2}, \qquad \bar{\kappa}_N=\kappa_\infty+\frac{b}{(N+1)^2}.
\end{equation}
This form applies to a spectral density that is smooth and positive inside its band and follows a power law at each band edge, such as that in Eq.~\eqref{chain_spectral_density}.
We approximate the leading corrections to the asymptotic chain coefficients as~\cite{kuijlaars2004riemann,chin2010exact}
\begin{equation}
\omega_{n+1}-\omega_\infty\simeq\frac{a}{n^2}, \qquad \kappa_n-\kappa_\infty\simeq\frac{b}{n^2},
\end{equation}
where $a$ and $b$ are obtained by least-squares fitting to the last third of the available coefficients $\{\omega_1,\dots,\omega_{N+1}\}$ and $\{\kappa_1,\dots,\kappa_N\}$, respectively.

Next, we propagate the self-energy toward the qubit by iterating Eq.~\eqref{eq:main BSE recurrence right} with decreasing site index down to $\Sigma_{2\to1}(p)$.
At the qubit boundary, the final step gives
\begin{equation}
\Sigma_{1\to0}(p)=\frac{g^2}{p+i\omega_1+\Sigma_{2\to1}(p)}.
\end{equation}
The Laplace-domain qubit excitation amplitude is
\begin{equation}\label{eq:qubit_amplitude_Laplace}
\mathcal A_0(p)=\frac{1}{p+i\epsilon+\Sigma_{1\to0}(p)}.
\end{equation}
Then, the qubit excitation amplitude in the time domain is obtained using the inverse Laplace transformation $A_0(t)=\mathcal L^{-1}\left[\mathcal A_0(p)\right]$, and the qubit population is
\begin{equation}\label{eq:qubit_population}
P_e(t)=|A_0(t)|^2.
\end{equation}

\section{Self-energy, spectral density, chain closures, and the cut-invariant distance}\label{sec:termination_methods}

In this section, we discuss the physical meaning of the self-energy and its relation to the spectral density, the semicircular spectral density of the BSE, the MC in terms of the self-energy, and a self-energy distance that is invariant under moving the system--environment cut.

\subsection{Self-energy and spectral density}\label{sec:self_energy_spectrum}

The spectral density $J(\omega)=\pi\sum_\mu|h_\mu|^2\delta(\omega-\Omega_\mu)$ encodes information about the environment~\cite{leggett1987dynamics,weiss2012quantum,breuer2002theory}, where $\Omega_\mu$ are the frequencies of the oscillators connected to the system in a star-like geometry, and $h_\mu$ their couplings to the qubit.
Eliminating the oscillators gives the exact self-energy at the system port~\cite{breuer2002theory,garraway1997nonperturbative,lambropoulos2000fundamental}, which contains the same information as the spectral density,
\begin{equation}\label{eq:B_star_integral}
\Sigma^{\text{exact}}_{1\to0}(p)=\sum_\mu\frac{|h_\mu|^2}{p+i\Omega_\mu} =\frac{1}{\pi}\int_{\omega_L}^{\omega_R}\frac{J(\nu)}{p+i\nu}d\nu,\quad \operatorname{Re}p>0,
\end{equation}
where $[\omega_L,\omega_R]$ is the smallest interval containing the support of $J(\omega)\geq0$.
The star and chain representations give the same self-energy at the system port, as shown in Appendix~\ref{app:star_chain_bse}.
The self-energy is analytic for $\operatorname{Re}p>0$ and satisfies $\operatorname{Re}\Sigma^{\mathrm{exact}}_{1\to0}(p)>0$ for nonzero system--environment coupling.
On the real-frequency axis, its real part recovers the spectral density and its imaginary part gives the frequency-dependent environment-induced shift~\cite{cohen2024atom},
\begin{equation}\label{eq:B_boundary_value}
\begin{split}
\Sigma^{\mathrm{exact}}_{1\to0}(0^+-i\omega)&=J(\omega)+i\Delta(\omega),
\\
\Delta(\omega)&=\frac{1}{\pi}\mathcal{P}\int_{\omega_L}^{\omega_R}\frac{J(\nu)}{\omega-\nu}d\nu.
\end{split}
\end{equation}
Here $\mathcal{P}$ denotes the Cauchy principal value, which removes the interval $(\omega-\varepsilon,\omega+\varepsilon)$ around the pole at $\nu=\omega$ and takes $\varepsilon\to0^+$.
Both parts follow from Eq.~\eqref{eq:B_star_integral} with the Sokhotski--Plemelj formula,
\begin{equation}\label{eq:Sokhotski--Plemelj formula}
\begin{split}
\frac{1}{0^++i(\nu-\omega)}=\pi\delta(\nu-\omega)-i\mathcal{P}\left[\frac{1}{\nu-\omega}\right],
\end{split}
\end{equation}
so $\Delta$ is the Hilbert transform of $J$.

The same decomposition holds at every cut.
In the chain representation of the environment, the inverse recurrence in Eq.~\eqref{eq:inverse BSE recurrence right} transforms the exact self-energy $\Sigma^{\mathrm{exact}}_{1\to0}(0^+-i\omega)$ to site $n$, so that
\begin{equation}
\Sigma^{\mathrm{exact}}_{n+1\to n}(0^+-i\omega)=J_n(\omega)+i\Delta_n(\omega),
\end{equation}
where $J_n(\omega)$ is the spectral density of the tail beyond site $n$ as seen from site $n$, with $J_0=J$ and $\kappa_0=g$.
For the initially vacuum bath, the spectral density and self-energy give the same correlation function,
\begin{equation}\label{eq:correlation_from_self_energy}
C(t)=\frac{1}{\pi}\int_{\omega_L}^{\omega_R}J(\omega)e^{-i\omega t}d\omega =\mathcal L^{-1}\left[\Sigma^{\mathrm{exact}}_{1\to0}(p)\right](t).
\end{equation}

\subsection{BSE and the Wigner semicircle}\label{sec:BSE_semicircle}

On the real-frequency axis, with $x=\omega-\omega_\infty$, the BSE in Eq.~\eqref{BSE_closure} is
\begin{equation}\label{eq:BSE_real_axis}
\Sigma_\infty(0^+-i\omega)=\frac{1}{2}\sqrt{4\kappa_\infty^2-x^2}+\frac{i}{2}x,\qquad|x|<2\kappa_\infty,
\end{equation}
while for $|x|>2\kappa_\infty$ it is purely imaginary.
Its real part, $J_\infty(\omega)=\frac{1}{2}\sqrt{4\kappa_\infty^2-x^2}$ inside the band and zero outside, is a spectral density supported on $[\omega_L,\omega_R]=[\omega_\infty-2\kappa_\infty,\omega_\infty+2\kappa_\infty]$~\cite{woods2014mappings}, which has the form of the Wigner semicircle.

The two parts have a simple meaning for the uniform tail.
For the Laplace-domain amplitudes with an initially vacuum chain, the equations of motion beyond site $n$ give $i\kappa_n\mathcal A_{n+1}=\Sigma_{n+1\to n}\mathcal A_n$.
An outgoing wave of excitation $\mathcal A_n\propto e^{ikn}$ on the uniform tail, with $\omega=\omega_\infty+2\kappa_\infty\cos k$ and $\sin k<0$, then gives $\Sigma_\infty=i\kappa_\infty e^{ik}$.
Its real part, $\kappa_\infty|\sin k|$, is half the group velocity $|d\omega/dk|$, so the tail absorbs at the rate at which the outgoing wave carries the excitation away.
Its imaginary part, $\kappa_\infty\cos k=x/2$, is the frequency shift produced by the band.
The absorption vanishes at the band edges, where the group velocity vanishes.
Outside the band, no wave propagates and only the shift remains, so the BSE closure does not absorb bound states.
The two tail parameters act on different parts of $\Sigma_\infty$.
At the band center, a small change $\delta\omega_\infty$ of the tail frequency changes $\Sigma_\infty$ by $-i\delta\omega_\infty/2$, a pure frequency shift, while a change $\delta\kappa_\infty$ of the coupling changes $\Sigma_\infty$ by $\delta\kappa_\infty$, a pure change of absorption.

The BSE is the Wigner semicircle written as a self-energy.
Indeed, $\Sigma_\infty(0^+-i\omega)=i\kappa_\infty^2G_{\rm sc}(x+i0^+)$, where $G_{\rm sc}(z)=(\pi\kappa_\infty^2)^{-1}\int J_\infty(\omega_\infty+y)(z-y)^{-1}dy$ is the resolvent of the semicircle, and the fixed-point equation~\eqref{eq:self referring} becomes
\begin{equation}\label{eq:semicircle_resolvent}
\kappa_\infty^2G_{\rm sc}^2(z)-zG_{\rm sc}(z)+1=0.
\end{equation}
Wigner obtained the semicircle from the moments of large random matrices~\cite{wigner1955characteristic,wigner1958distribution}, and Pastur from the self-consistent equation~\eqref{eq:semicircle_resolvent} for the resolvent~\cite{pastur1972spectrum}.
In the BSE construction, the semicircle follows from the self-closing recurrence of the self-energy in the asymptotic region (Eq.~\eqref{eq:self referring}).

\subsection{Markovian closure in self-energy form}\label{sec:MC_self_energy}

Based on the spectral density, several approximation methods have been proposed.
The Born--Markov approximation uses time-local (memoryless) dissipative terms and frequency shifts~\cite{breuer2002theory}.
The pseudomode method uses one or multiple modes with Markovian damping~\cite{garraway1997nonperturbative,muller2026one}.
A related realization is a finite chain with site-dependent damping rates~\cite{sanchez2021accurate}.
Here we discuss the MC~\cite{nusseler2022fingerprint}, which terminates a chain using interconnected damped modes.
Its main purpose is to reduce computational cost while maintaining simulation accuracy.

The MC keeps the exact chain up to site $N$ and closes it with $M$ damped modes.
The closure modes, with creation (annihilation) operators $d_j^\dagger$ ($d_j$), are coupled to their nearest neighbors, and each is also coupled to site $N$,
\begin{equation}\label{eq:H_MC}
\begin{split}
H_\mathrm{MC}=&\sum_{j=1}^{M}\tilde\omega_j d_j^\dagger d_j+\sum_{j=1}^{M-1}\tilde\kappa_j\big(d_j^\dagger d_{j+1}+d_{j+1}^\dagger d_j\big)
\\
&+\sum_{j=1}^{M}\big(c_j a_N^\dagger d_j+c_j^* d_j^\dagger a_N\big),
\end{split}
\end{equation}
where each closure mode is damped by the Lindblad dissipator
\begin{equation}
\mathcal{D}[\rho]=\sum_{j=1}^{M}\Gamma_j\Big(d_j\rho d_j^\dagger-\frac{1}{2}\big\{d_j^\dagger d_j,\rho\big\}\Big).
\label{eq:D_MC}
\end{equation}
As a damped harmonic environment coupled linearly to site $N$, the closure presents the following self-energy to that site,
\begin{equation}
\Sigma^\mathrm{MC}_{N+1\to N}(p)=\mathbf{c}^T\Big(pI_M+i\tilde H+\frac{1}{2}\tilde\Gamma\Big)^{-1}\mathbf{c}^\ast,
\label{eq:Sigma_MC}
\end{equation}
where $\tilde H$ is the tridiagonal matrix with the closure-mode frequencies $\tilde\omega_j$ on the diagonal and the couplings $\tilde\kappa_j$ between neighboring closure modes off the diagonal, $\tilde\Gamma=\mathrm{diag}(\Gamma_1,\dots,\Gamma_M)$, and $\mathbf{c}=(c_1,\dots,c_M)^T$.
The parameters are fitted once for each closure size $M$ to the time-domain correlation function $C_\infty(t)$ of the Wigner semicircle residual spectral density, which carries the same information as the BSE $\Sigma_\infty$,
\begin{equation}\label{eq:correlation_for_MC}
C_\infty(t)=\frac{1}{\pi}\int_{\omega_L}^{\omega_R}J_\infty(\omega)e^{-i\omega t}d\omega =\mathcal L^{-1}\left[\Sigma_\infty(p)\right](t),
\end{equation}
then shifted and rescaled for the bath under consideration.
The correlation function and the self-energy describe the same response, but different time- and frequency-domain fitting objectives need not select the same parameters.

In our implementation, the complexity of simulating dynamics with $\Sigma_\infty$ is close to that with the spectral density $J(\omega)$, since both require continuum integration.
In cases where computational efficiency is key, a finite set of damped harmonic oscillators fitted to the BSE, such as the MC, provides a more efficient representation.
To compare such closures with the BSE closure, and to fit them to their corresponding targets, we use a cut-invariant error measure.

\subsection{Cut-invariant distance}\label{sec:cut_invariant_error}
In this subsection, we give the definition of a distance between two environmental responses (self-energies) that is invariant under shifts of the system--environment cut.
Let $\Sigma_a$ and $\Sigma_b$ be two self-energies presented to the same site, for example a closure and the exact tail.
At frequencies where the boundary values are finite and their real parts are positive, we define
\begin{equation}\label{eq:cut_invariant_error}
\chi(\omega)=\frac{|\Sigma_a(p)-\Sigma_b(p)|}{|\Sigma_a(p)+\Sigma_b^*(p)|},\qquad p=0^+-i\omega,
\end{equation}
with $0\leq\chi\leq1$, and $\chi$ is the pseudo-hyperbolic distance on the right half-plane~\cite{garnett2007bounded}.

On the real-frequency axis, the recurrence maps in Eqs.~\eqref{eq:main BSE recurrence left},~\eqref{eq:inverse BSE recurrence left},~\eqref{eq:main BSE recurrence right}, and~\eqref{eq:inverse BSE recurrence right} leave $\chi$ unchanged for a lossless step with nonzero coupling.
Hence, if the self-energies of two closures attached at the same cut are propagated through the same lossless chain segment with nonzero couplings, $\chi$ has the same value at every site from the cut to the system port.
We call the self-energy distance $\chi$ on the real-frequency axis the cut-invariant distance.
When $\chi$ is used as an error measure, it is called the cut-invariant error.

For $p=\eta-i\omega$ with $\eta>0$, denote the self-energy distance by $\chi_\eta$ and one recurrence step by $F_{n,p}(\Sigma)=\kappa_n^2/(p+i\omega_{n+1}+\Sigma)$.
Then
\begin{equation}\label{eq:chi_contraction}
\chi_\eta(F_{n,p}(\Sigma_a),F_{n,p}(\Sigma_b))=\frac{|\Sigma_a-\Sigma_b|}{|\Sigma_a+\Sigma_b^*+2\eta|}\leq\chi_\eta(\Sigma_a,\Sigma_b).
\end{equation}
For fixed $\omega$ and $\eta>0$, the distance $\chi_\eta$ between the two self-energies at the cut is an upper bound on their distance at the system port after propagation through the same chain segment.
Moving across a lossy part of the chain also gives a contraction.

The cut-invariant error $\chi(\omega)$ alone does not determine the error in the system dynamics over a finite observation interval.
For example, a hard cutoff has $\chi=1$ at every depth, even when the reflected excitation does not return within the observation interval.

Next we discuss the error budget, which separates the closure error from the depth error.
Consistent with Eq.~\eqref{time-nonlocal_bse_p}, we define the asymptotic closure target at site $N$ by
\begin{equation}\label{eq:local_BSE_target}
\Sigma_{N+1\to N}^{\rm target}(p)=\frac{\kappa_N^2}{p+i\omega_{N+1}+\Sigma_\infty(p)}.
\end{equation}
This target requires the local coefficient $\omega_{N+1}$ in addition to the retained chain and the attachment coupling $\kappa_N$.
For the MC, the closure error $\chi_M=\chi(\Sigma^{\rm MC}_{N+1\to N},\Sigma_{N+1\to N}^{\rm target})$ measures how closely the MC reproduces the target self-energy at site $N$.
The depth error $\chi_N=\chi(\Sigma_{N+1\to N}^{\rm target},\Sigma^{\rm exact}_{N+1\to N})$ measures how closely that target approximates the exact tail self-energy at the same site.
With the total cut-invariant error
\begin{equation}\label{eq:total cut-invariant error}
 \chi_{\rm tot}=\chi(\Sigma^{\rm MC}_{N+1\to N},\Sigma^{\rm exact}_{N+1\to N}), 
\end{equation}
 the hyperbolic triangle and reverse-triangle inequalities give~\cite{garnett2007bounded}
\begin{equation}\label{eq:chi_budget}
\frac{|\chi_M-\chi_N|}{1-\chi_M\chi_N}\leq\chi_{\rm tot}\leq\frac{\chi_M+\chi_N}{1+\chi_M\chi_N}.
\end{equation}
These bounds hold at each frequency $\omega$ for which the three self-energies $\{\Sigma^{\rm MC}_{N+1\to N},\Sigma_{N+1\to N}^{\rm target},\Sigma^{\rm exact}_{N+1\to N}\}$ are finite and have positive real parts.
The closure error and the depth error can reinforce or cancel.
When one is much smaller than the other, the total error is dominated by the larger one.
The same bounds hold for the chain-extended BSE when $\bar{\Sigma}_N$ replaces $\Sigma_\infty$ in Eq.~\eqref{eq:local_BSE_target}.
If either analytic BSE target is used directly as the closure, the closure error is zero and $\chi_{\rm tot}=\chi_N$.
To use $\chi$ for numerical selection of a closure, the frequency interval, frequency weights, and any finite broadening must also be specified.

\section{Numerical demonstration}\label{Numerical_demonstration}

This section compares the BSE closures with finite MC realizations in a numerical benchmark.
Frequencies and couplings are expressed in units of the upper cutoff frequency $\omega_c$, and time in units of $\omega_c^{-1}$, with $\omega_c=1$.
The spectral band is $[\omega_L,\omega_R]=[0,\omega_c]$.
As a benchmark, we use a structured Ohmic environment with the spectral density
\begin{equation}\label{chain_spectral_density}
J(\omega)=\frac{2\pi\alpha}{\mathcal N}\omega\left[1+Ae^{-(\omega-\omega_0)^2/(2\sigma^2)}\right], \quad 0\leq\omega\leq\omega_c,
\end{equation}
where $\alpha$ sets the system--environment coupling strength.
The normalization $\mathcal N$ fixes $g^2=\alpha=\pi^{-1}\int_0^1J(\omega)d\omega$ (see Appendix~\ref{app:star_chain_bse}).
Here we take $\alpha=0.02$, $\epsilon=0.5$, $A=4$, $\omega_0=0.70$, and $\sigma=0.07$, giving $g=\sqrt{0.02}\simeq0.1414$.
The construction of the chain coefficients $\omega_n$ and $\kappa_n$ is given in Appendix~\ref{app:star_chain_bse}.
In the asymptotic region,
\begin{equation}
\omega_n\to\omega_\infty=\frac{\omega_c}{2}, \qquad \kappa_n\to\kappa_\infty=\frac{\omega_c}{4}.
\end{equation}
The reference dynamics are obtained numerically from the exact continuum spectral density $J(\omega)$ in Eq.~\eqref{chain_spectral_density}, and the observation interval is $0\leq t\leq T$, with $T=2000$.
Across all cases, $N$ denotes the number of explicit chain modes, i.e., the retained chain sites $1,\dots,N$ that are treated exactly, while the chain beyond site $N$ is replaced by a closure.

We compare five closures attached at site $N$:
\begin{enumerate}
\item the asymptotic BSE closure, given by Eq.~\eqref{time-nonlocal_bse_p};
\item the chain-extended BSE closure, which replaces $\Sigma_\infty$ in Eq.~\eqref{time-nonlocal_bse_p} by $\bar{\Sigma}_N$ from Eq.~\eqref{Extended_BSE_closure};
\item the semicircle-fit MC, fitted to the time-domain correlation function of the Wigner semicircle as in Ref.~\cite{nusseler2022fingerprint}, with $M=10$ closure modes;
\item the $\chi$ fit to $\bar{\Sigma}_N$, an MC fitted to the chain-extended BSE closure using the cut-invariant error $\chi$, with $M=10$;
\item the $\chi$ fit to the exact tail, an MC fitted to the exact tail self-energy $\Sigma^{\rm exact}_{N+1\to N}$ using $\chi$ and $J(\omega)$, with $M=10$.
The exact tail is obtained by using the inverse recurrence in Eq.~\eqref{eq:inverse BSE recurrence right} to propagate the exact self-energy $\Sigma^{\rm exact}_{1\to 0}$ to site $N$.
\end{enumerate}
The parameters of the two $\chi$ fits are selected at each $N$ by minimizing the mean of $\chi^2$ on the real-frequency axis with uniform frequency weights.
The $\chi$ fit to $\bar{\Sigma}_N$ uses $1201$ equally spaced frequencies over $|\omega-\bar{\omega}_N|\leq1.96\bar{\kappa}_N$.
The $\chi$ fit to the exact tail uses $2001$ equally spaced frequencies over $0.01\leq\omega\leq0.99$.
The oscillator frequencies, nearest-neighbor couplings, attachment couplings, and nonnegative damping rates are varied.
Both fits also include a weak penalty on their deviation from the chain-extended BSE closure outside its band.
Using $x=(\omega-\bar{\omega}_N)/\bar{\kappa}_N$, this penalty samples $150$ frequencies on each interval $2.05\leq|x|\leq8$ and weights the complex residual by $0.3$, with both responses expressed in units of $\kappa_N^2/\bar{\kappa}_N$.
The squared residuals inside and outside the band are summed and divided by the number of in-band fitting points.
The in-band fitting windows exclude the band edges, where $\chi$ can approach $1$ even for accurate closures.

\begin{figure}[t]
\centering
\includegraphics[width=\columnwidth]{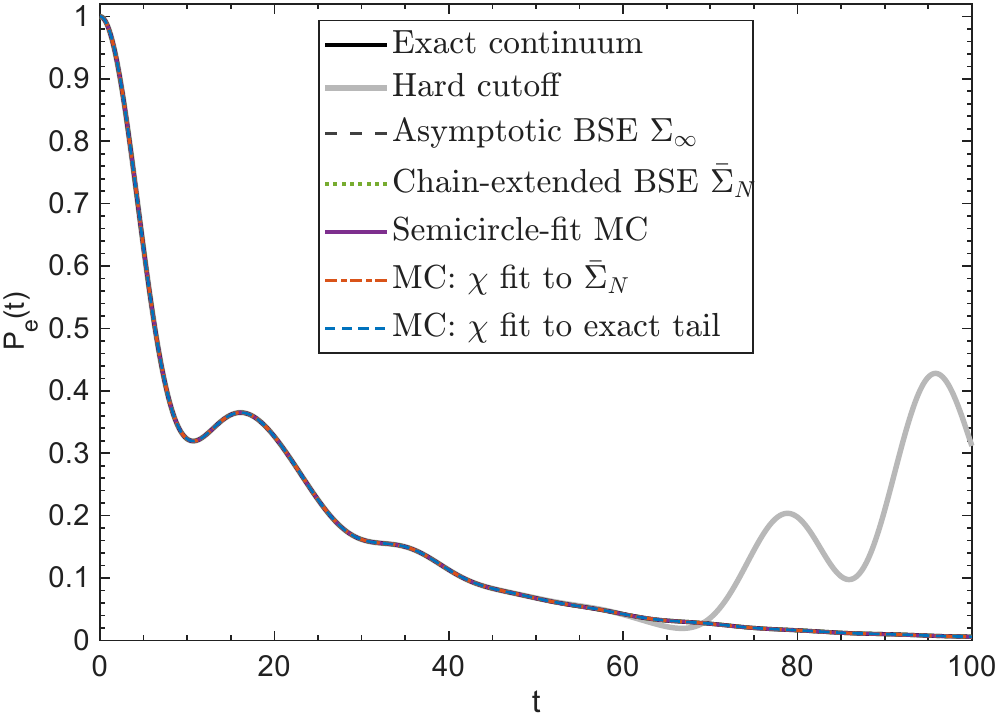}
\caption{
Qubit excited-state population $P_e(t)$ for the hard cutoff and the five closures with $N=15$ explicit chain modes, compared with the exact continuum reference.
The displayed interval is $0\leq t\leq100$ and the maximum observation time is $T=2000$.
}
\label{fig:time_domain}
\end{figure}

Figure~\ref{fig:time_domain} compares the time evolution of the qubit population $P_e(t)$ for a hard cutoff and the five closures, with $N=15$.
The reference is the exact continuum obtained by using the spectral density.
The hard cutoff produces an unphysical reflection after the round-trip time $2N/v\simeq60$, where $v=2\kappa_\infty$ is the group velocity at $\epsilon=\omega_\infty$.
All five closures closely follow the reference over the displayed time interval.

\begin{figure}[t]
\centering
\includegraphics[width=\columnwidth]{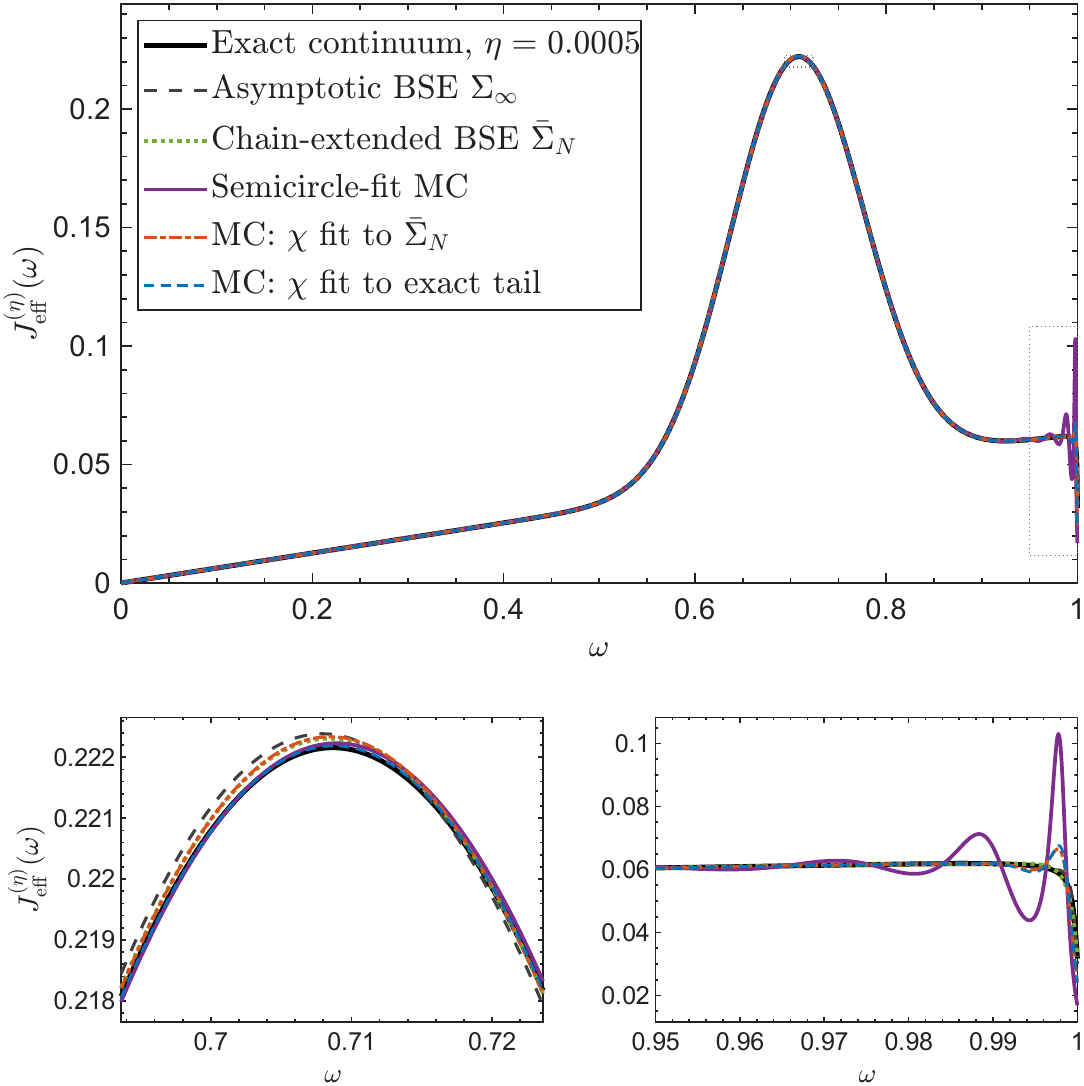}
\caption{
Broadened effective spectral densities at the system port for $N=15$ and $\eta=0.0005$, compared with the exact continuum reference at the same broadening.
The lower panels enlarge the peak and the upper band edge, marked by the dotted boxes in the upper panel.
}
\label{fig:effective_spectrum}
\end{figure}

Figure~\ref{fig:effective_spectrum} compares the effective spectral density seen by the qubit for the five closures, with $N=15$.
We use the broadened effective spectral density obtained from the self-energy at the system port,
\begin{equation}
J_{{\rm eff},X}^{(\eta)}(\omega)=\operatorname{Re}\Sigma^{X}_{1\to0}(\eta-i\omega),
\end{equation}
where $\eta=0.0005$ and $X$ denotes the closure.
The reference is the broadened exact continuum,
\begin{equation}
\begin{aligned}
J_{\rm ref}^{(\eta)}(\omega) &=\operatorname{Re}\Sigma^\mathrm{exact}_{1\to0}(\eta-i\omega)
\\
&=\frac{1}{\pi}\int_0^1\frac{\eta J(\nu)} {(\omega-\nu)^2+\eta^2}d\nu.
\end{aligned}
\end{equation}
It approaches $J(\omega)$ as $\eta\to0^+$ for $0<\omega<1$.
At the peak, all closures closely agree with the reference.
Near the upper band edge, the semicircle-fit MC oscillates around the reference from $\omega\simeq0.98$, the two $\chi$-fitted MCs deviate for $\omega\gtrsim0.995$, and the two BSE closures follow the reference up to $\omega\simeq0.999$.

\begin{figure*}[t]
\centering
\figpanel{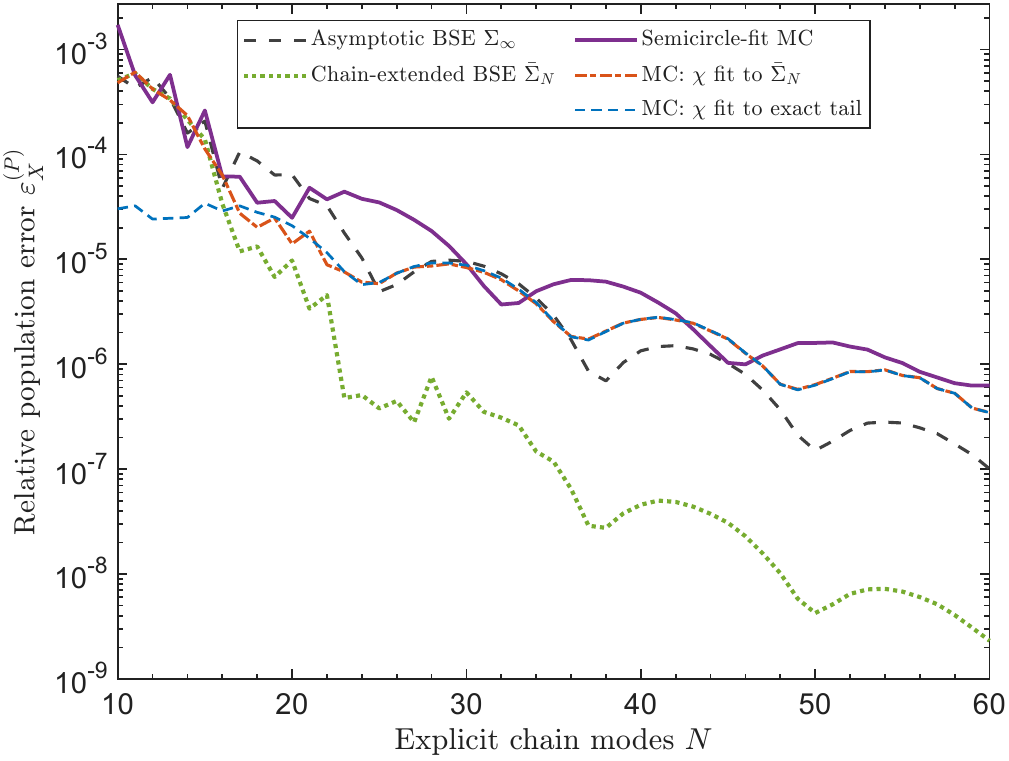}{a}\hfill
\figpanel{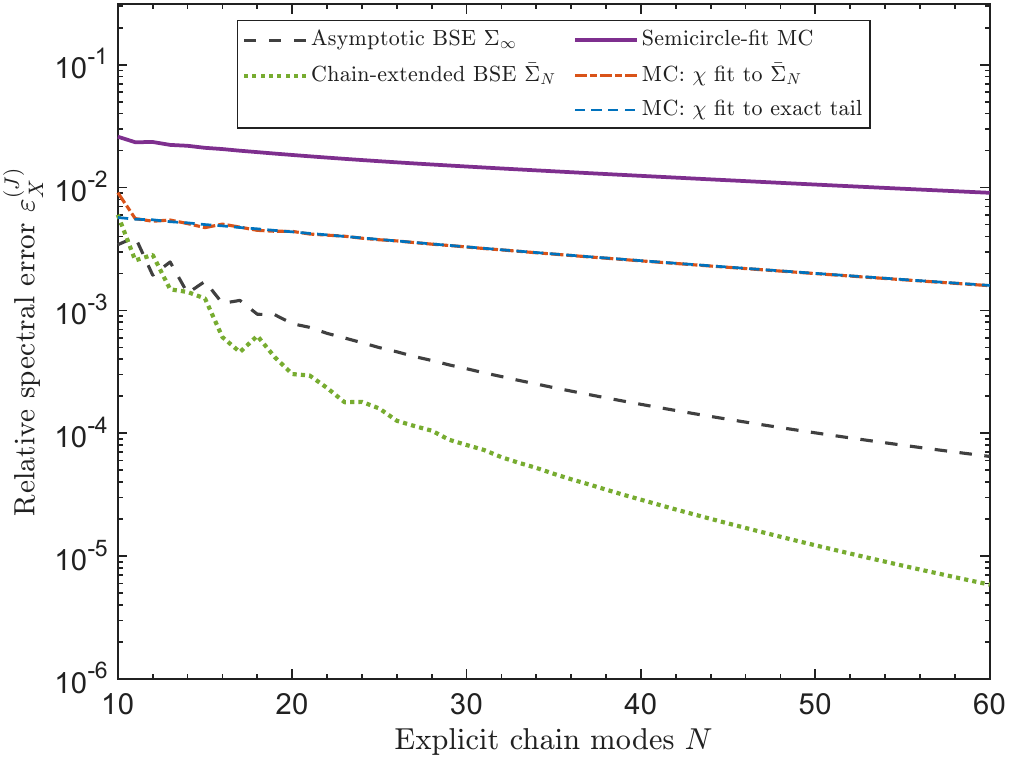}{b}
\\[1ex]
\figpanel{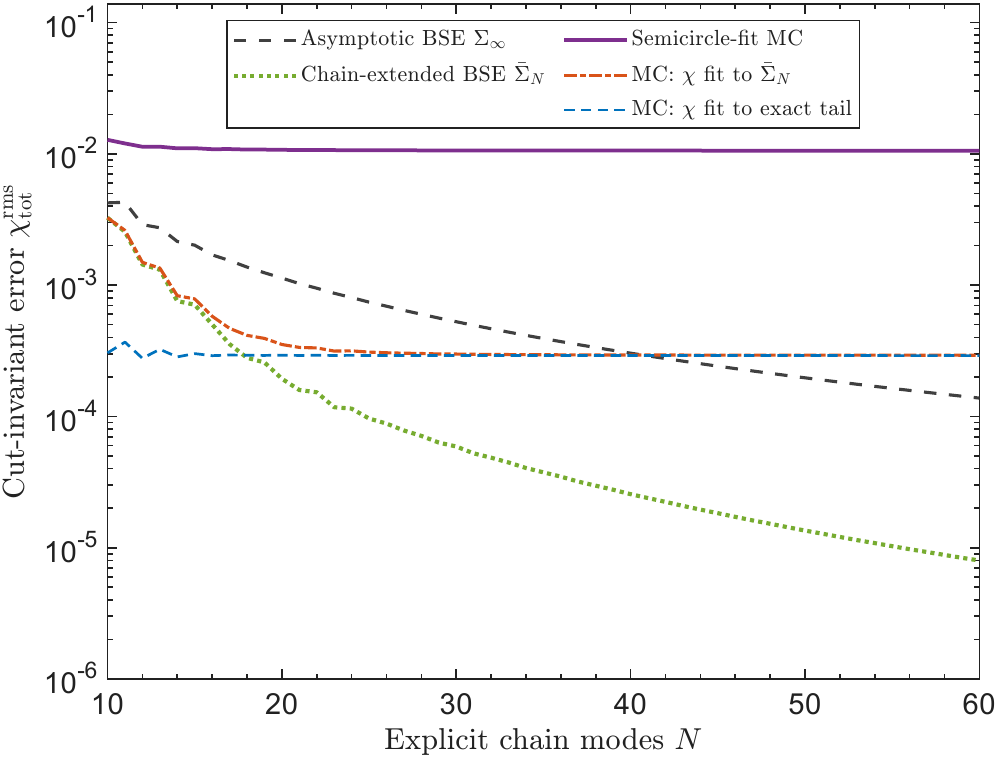}{c}\hfill
\figpanel{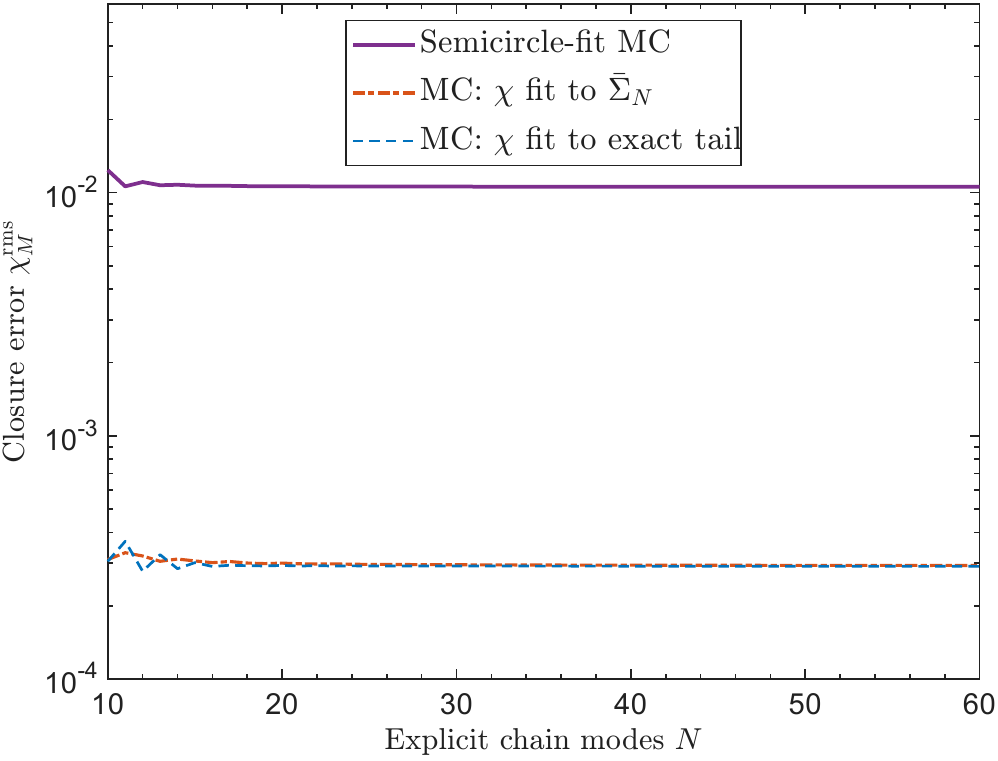}{d}
\caption{
Errors of the closures as a function of the number of explicit chain modes $N$.
(a) Relative population error $\varepsilon_X^{(P)}(N;T)$, defined in Eq.~\eqref{eq:population_error}, with $T=2000$, evaluated on the time grid with $\Delta t=0.1$.
(b) Relative spectral error $\varepsilon_X^{(J)}(N;\eta)$, defined in Eq.~\eqref{eq:spectral_error}, with the common broadening $\eta=0.0005$.
(c) Total cut-invariant error $\chi_{\rm tot}^{\rm rms}$ against the exact tail, which for the BSE closures is the depth error.
(d) Closure error $\chi_M^{\rm rms}$ of the three MCs with $M=10$ against their comparison targets: the asymptotic BSE closure, the chain-extended BSE closure, and the exact tail, respectively.
The integrals in (b) are taken over the full band $0\leq\omega\leq1$, and the root mean squares in (c) and (d) over the band interior $0.01\leq\omega\leq0.99$.
}
\label{fig:errors_vs_N}
\end{figure*}

To quantify the accuracy of each closure in the time domain, we define the relative population error
\begin{equation}\label{eq:population_error}
\varepsilon_X^{(P)}(N;T)=\left[\frac{\int_0^T\left|P_e^{(X)}(t)-P_e^{\rm ref}(t)\right|^2dt}{\int_0^T\left|P_e^{\rm ref}(t)\right|^2dt}\right]^{1/2},
\end{equation}
where $P_e^{\rm ref}(t)$ is the exact continuum reference population.
Figure~\ref{fig:errors_vs_N}(a) shows $\varepsilon_X^{(P)}$ for $N=10,\dots,60$.
The $\chi$ fit to the exact tail gives the smallest error for $N\leq16$.
Its target is the exact tail and has no depth error, whereas the other closures rely on the asymptotic or chain-extended BSE, whose depth errors are still large at small $N$ (Fig.~\ref{fig:errors_vs_N}(c)).
For $N\geq17$, the chain-extended BSE gives the smallest error, about $40$ times smaller than the next best closure at $N=60$.
For $N\geq47$, the semicircle-fit MC gives the largest error.
The errors do not decrease monotonically with $N$, because an imperfect closure can reflect part of the excitation back toward the qubit.
Writing $A_0^{(X)}=A_0^{\rm ref}+\delta A_0$, the population difference is $P_e^{(X)}-P_e^{\rm ref}=2\operatorname{Re}[(A_0^{\rm ref})^*\delta A_0]+|\delta A_0|^2$.
The cross term depends on the phase of $\delta A_0$ relative to $A_0^{\rm ref}$.
Each added site changes the return time and phase of the reflected excitation, which can contribute to the observed oscillations of $\varepsilon_X^{(P)}$ with $N$, with a period of about $12$ sites.

\begin{figure*}[t]
\centering
\figpanel{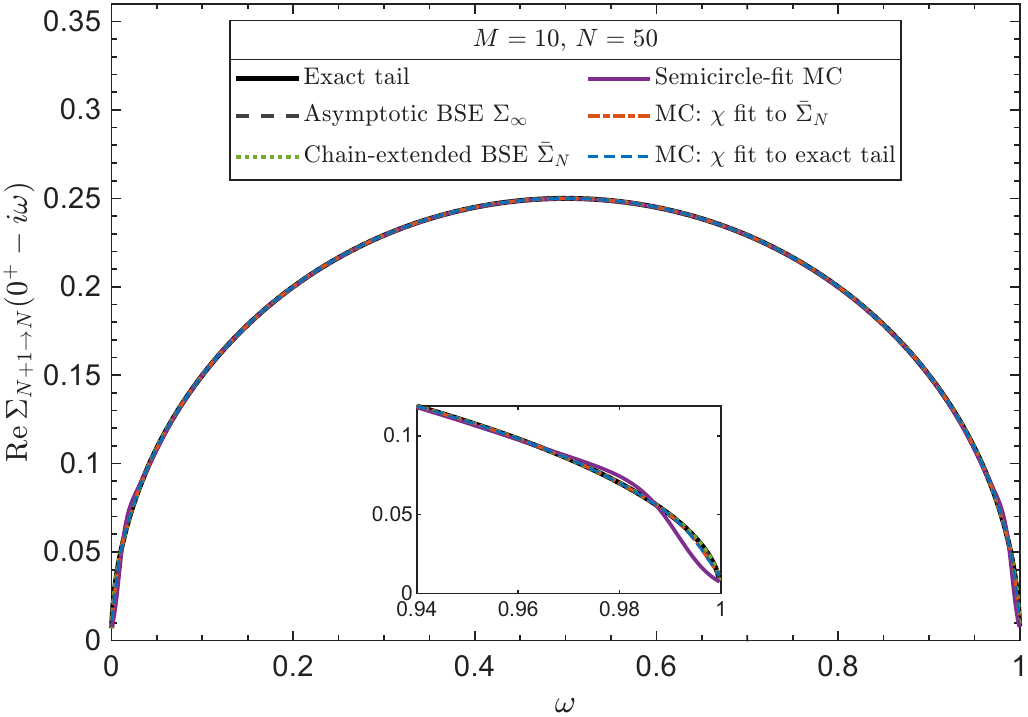}{a}\hfill
\figpanel[0.13,0.95]{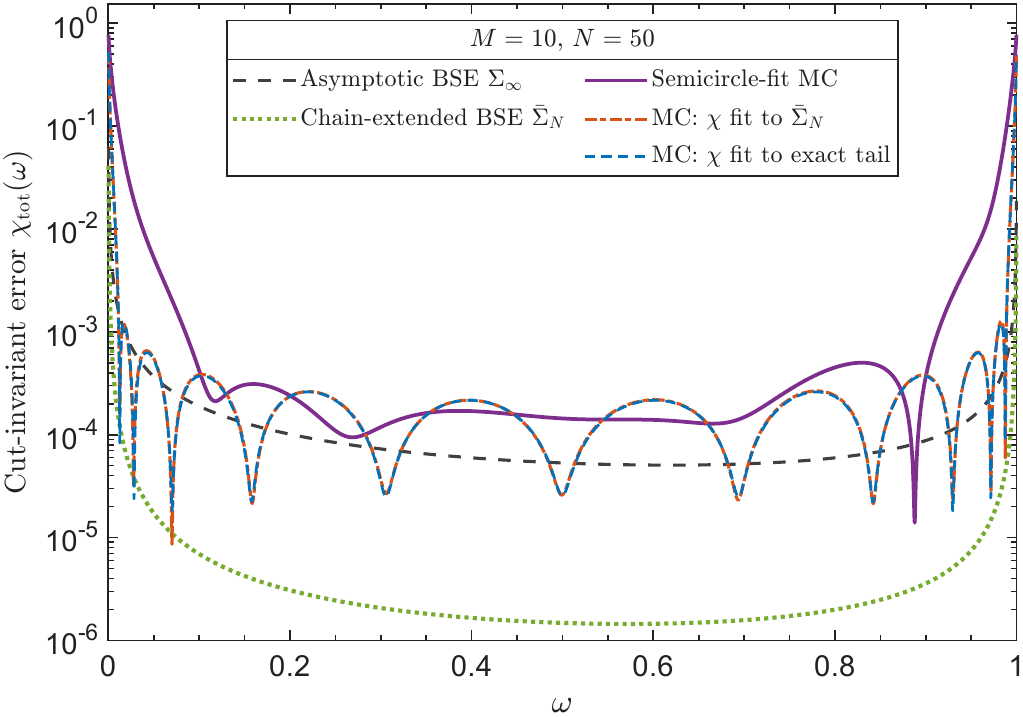}{b}
\caption{
Self-energies and cut-invariant errors at site $N=50$ with $M=10$.
(a) Real parts of the self-energies presented to site $N$ by the three MCs and by their comparison targets, with the upper band edge enlarged in the inset.
(b) Total cut-invariant error $\chi_{\rm tot}(\omega)$ of the five closures against the exact tail.
}
\label{fig:cut_N50}
\end{figure*}

In addition to the time-domain population error, we quantify the accuracy of the closures using the relative spectral error
\begin{equation}\label{eq:spectral_error}
\varepsilon_X^{(J)}(N;\eta)= \left[\frac{\int_0^1\left|J_{{\rm eff},X}^{(\eta)}(\omega)-J_{\rm ref}^{(\eta)}(\omega)\right|^2d\omega}{\int_0^1\left|J_{\rm ref}^{(\eta)}(\omega)\right|^2d\omega}\right]^{1/2}.
\end{equation}
All spectra use the same broadening $\eta=0.0005$.
Figure~\ref{fig:errors_vs_N}(b) presents the relative spectral error as a function of $N$.
The spectral errors decrease overall with $N$, and a BSE closure gives the smallest error at every tested $N$, with the chain-extended BSE smallest for $N\geq15$.
The errors of the three MCs decrease only slowly, to $9.1\times10^{-3}$ for the semicircle-fit MC and $1.6\times10^{-3}$ for the two $\chi$ fits at $N=60$.

In Figs.~\ref{fig:errors_vs_N}(c), \ref{fig:errors_vs_N}(d), and~\ref{fig:cut_N50}, the broadening used in all the cut-invariant errors and the real parts of the self-energies is set to be $\eta^\prime=0^+$, so they are evaluated on the real-frequency axis.
Figure~\ref{fig:errors_vs_N}(c) shows the total cut-invariant error $\chi_{\rm tot}^{\rm rms}$ of the five closures against the exact tail at site $N$.
Since $\chi$ can approach $1$ near a band edge despite a small absolute mismatch, we report the root mean square $\chi^{\rm rms}$ over the common band interior $0.01\leq\omega\leq0.99$,
\begin{equation}\label{eq:chi_rms}
\chi^{\rm rms}=\left[\frac{1}{0.98}\int_{0.01}^{0.99}\chi^2(\omega)d\omega\right]^{1/2}.
\end{equation}
The spectral error in Eq.~\eqref{eq:spectral_error} stays finite at the band edges, so we integrate it over the full band.

For the two BSE closures, $\chi_{\rm tot}^{\rm rms}$ is the depth error and decreases overall with $N$.
The errors of the three MCs level off, at $1.1\times10^{-2}$ for the semicircle-fit MC and at about $2.9\times10^{-4}$ for the two $\chi$ fits.
The $\chi$ fit to $\bar{\Sigma}_N$ follows the chain-extended BSE up to $N\simeq15$ before it levels off, and for $N\geq18$ the chain-extended BSE gives the smallest error.
For $N\geq41$, the asymptotic BSE also falls below the two $\chi$ fits.

To explain this behavior, Fig.~\ref{fig:errors_vs_N}(d) shows the closure errors of the three MCs against their corresponding targets.
For the semicircle-fit MC, the closure error levels off at $1.1\times10^{-2}$, which is the error $\chi^{\rm rms}$ of its universal fit to $C_\infty(t)$ in Eq.~\eqref{eq:correlation_for_MC}, measured against $\Sigma_\infty$.
The two $\chi$ fits reduce the closure error to about $2.9\times10^{-4}$.
A $\chi$ fit of the MC to $\Sigma_\infty$ gives $\chi^{\rm rms}\simeq3.1\times10^{-3}$, $5.9\times10^{-4}$, and $2.9\times10^{-4}$ for $M=6$, $8$, and $10$, and the value for $M=10$ equals the closure error of the two $\chi$ fits at large $N$.
As the depth error becomes small compared with the closure error, the latter dominates the total error, consistent with Eq.~\eqref{eq:chi_budget} and the plateaus in Fig.~\ref{fig:errors_vs_N}(c).
The total errors of the two $\chi$ fits approach each other and are almost identical at large $N$.

Figure~\ref{fig:cut_N50}(a) compares the real parts of the self-energies of the three MCs and their comparison targets at site $N=50$, with the exact tail self-energy $\Sigma^{\rm exact}_{N+1\to N}$ as the reference.
All responses are close to the reference, and the visible differences appear near the band edges.

Figure~\ref{fig:cut_N50}(b) shows the frequency dependence of $\chi_{\rm tot}$ at $N=50$.
The chain-extended BSE gives the smallest error over almost the whole band interior.
The error of the asymptotic BSE varies smoothly, with a minimum of about $5\times10^{-5}$.
The two $\chi$ fits nearly coincide, oscillate between about $2\times10^{-5}$ and $4\times10^{-4}$ for $0.1\leq\omega\leq0.9$, and lie above the asymptotic BSE over most of the band.
The error of the semicircle-fit MC follows its closure error and stays below about $5\times10^{-4}$ for $0.1\leq\omega\leq0.9$.
The BSE closures remain accurate close to the band edges, where the MCs deviate most.

\section{Discussion}\label{Discusses}

For a harmonic bath linearly coupled to the system, initially in vacuum and uncorrelated with the system, the correlation function determines its response to the system~\cite{FEYNMAN1963118,tamascelli2018nonperturbative}.
A self-energy can be converted into a correlation function and supplied to existing methods for nonlinear or multi-excitation dynamics.
If the cut is moved through an initially correlated bath, the initial correlations between the explicit chain modes and the tail must also be considered.

In the numerical benchmark, the chain-extended BSE gives the smallest population, spectral, and cut-invariant errors for $N\geq18$, though it requires chain coefficients that follow the assumed trend.
Fitting the MC with the cut-invariant error lowers its closure error for $M=10$ from $1.1\times10^{-2}$ to $2.9\times10^{-4}$, and the fit to the chain-extended BSE reaches the accuracy of the fit to the exact tail at large $N$.

The BSE closures can be particularly useful when an explicit spectral density is unavailable, but the chain coefficients and information about the asymptotic tail are known.
Examples include baths specified directly by a microscopic Hamiltonian~\cite{haydock1972electronic,papaefstathiou2025efficient}, structured photonic waveguides and coupled-resonator arrays~\cite{sundaresan2019interacting}, circuit environments~\cite{bakr2026exact}, and environments with limited structural information~\cite{viswanath1994recursion,gautschi2004orthogonal}.
In these cases, the closure provides an approximation to the tail based on its asymptotic behavior.

\section{Conclusion}\label{Conclusion}

Starting from the system--environment cut, we identify the self-energy as the environmental response presented to the system, with its recurrence forming a covariant transformation under shifts of the system--environment cut.
Under this transformation law, the self-energy reaches a fixed point in the asymptotic region, which is the BSE.
The chain-extended BSE further exploits the trend of the retained chain to obtain higher accuracy.
The self-energy also defines a cut-invariant distance, whose value at the cut equals its value at the qubit port on the real-frequency axis, and which separates the depth error from the closure error of a finite realization.

For the BSE closures and the MC fits to the BSE, once the available chain coefficients and the asymptotic tail behavior are known, an explicit spectral density is not required.
This could provide a basis for the inverse design of structured environments.
Natural next developments include tests of multi-excitation dynamics, operator-valued BSEs for more general structures, and applications to continuous-variable systems.

\begin{acknowledgments}
This project uses ChatGPT, Claude, and MATLAB for coding and plotting.
All results reported here were reviewed and verified by the authors.
\end{acknowledgments}

\section*{Data availability}
The data supporting this research are publicly available in the GitHub repository at \url{https://github.com/RaymondtheSolowing/Self-Energy-Construction-for-Chain-Representations-of-Open-Quantum-Systems}.

\appendix
\section{Green-function representation and boundary self-energy recurrence}
\label{app:bse_recurrence}

The Hamiltonian in Eq.~\eqref{eq:H_F} conserves excitation number and satisfies $H|\mathrm{vac}\rangle=0$, so $i\dot B(t)=0$ and the vacuum amplitude decouples from the excitation amplitudes.
In the single-excitation sector, define the resolvent and its matrix elements by
\begin{equation}
\mathbf G(p)=(pI+i\mathcal J_{\mathrm H})^{-1},\qquad G_{mn}(p)=\langle m|\mathbf G(p)|n\rangle.
\end{equation}
For the initially excited qubit and vacuum chain, $\mathcal A_0(p)=G_{00}(p)$.

In this chain, an interior site $n$ has two boundary ports connecting it to the remaining degrees of freedom, one through site $n-1$ and one through site $n+1$.
For the $n-1\to n$ port with $n\geq 2$, the self-energy is
\begin{equation}\label{eq: Schur BSE left}
\begin{split}
&\Sigma_{n-1\to n}(p)
\\
&=\kappa_{n-1}^2\left\langle n-1\left|\left(pI_{0:n-1}+i\mathcal{J}_{0:n-1}\right)^{-1}\right|n-1\right\rangle,
\end{split}
\end{equation}
where $\mathcal{J}_{0:n-1}$ is the Jacobi matrix of the part from the qubit through site $n-1$.
Using $\kappa_0=g$, we write $pI_{0:n-1}+i\mathcal{J}_{0:n-1}$ in block form as
\begin{equation}\label{block_chain}
\begin{split}
&pI_{0:n-1}+i\mathcal{J}_{0:n-1}
\\
=&
\begin{pmatrix}
pI_{0:n-2}+i\mathcal{J}_{0:n-2} & i\kappa_{n-2}(0,\ldots,1)^T \\
i\kappa_{n-2}(0,\ldots,1) & p+i\omega_{n-1}
\end{pmatrix}.
\end{split}
\end{equation}
For a block matrix $M= \begin{pmatrix} A & B \\ C & D \end{pmatrix}$, the Schur-complement formula gives $(D-CA^{-1}B)^{-1}$ for the lower-right block of $M^{-1}$.
Identifying the blocks in Eq.~\eqref{block_chain}, we have
\begin{equation}\begin{split}
&\langle n-1|\left(pI_{0:n-1}+i\mathcal{J}_{0:n-1}\right)^{-1}|n-1\rangle
\\
=& \big(p+i\omega_{n-1}
\\
&+\kappa_{n-2}^2\langle n-2|\left(pI_{0:n-2}+i\mathcal{J}_{0:n-2}\right)^{-1}|n-2\rangle\big)^{-1}.
\end{split}\end{equation}
Using the self-energy definition at the preceding port,
\begin{equation}
\Sigma_{n-2\to n-1}(p)= \kappa_{n-2}^2\langle n-2|\left(pI_{0:n-2}+i\mathcal{J}_{0:n-2}\right)^{-1}|n-2\rangle,
\end{equation}
Eq.~\eqref{eq: Schur BSE left} becomes the recurrence relation
\begin{equation}\label{eq:BSE recurrence left}
\Sigma_{n-1\to n}(p)= \kappa_{n-1}^2\left[p+i\omega_{n-1}+\Sigma_{n-2\to n-1}(p)\right]^{-1}.
\end{equation}

Similarly, for the $n+1\to n$ port, the self-energy is
\begin{equation}\label{eq: Schur BSE right}
\begin{split}
&\Sigma_{n+1\to n}(p)
\\
&=\kappa_n^2\left\langle n+1\left|\left(pI_{n+1:\infty}+i\mathcal{J}_{n+1:\infty}\right)^{-1}\right|n+1\right\rangle,
\end{split}
\end{equation}
where $\mathcal{J}_{n+1:\infty}$ is the Jacobi matrix of the semi-infinite structure beginning at site $n+1$.
In block form,
\begin{equation}
\begin{split}
&pI_{n+1:\infty}+i\mathcal{J}_{n+1:\infty}
\\
=&
\begin{pmatrix}
p+i\omega_{n+1} & i\kappa_{n+1}(1,0,\ldots) \\
i\kappa_{n+1}(1,0,\ldots)^T & pI_{n+2:\infty}+i\mathcal{J}_{n+2:\infty}
\end{pmatrix}.
\end{split}
\end{equation}
Here the lower-right entry is an infinite-dimensional block representing sites $(n+2,n+3,\ldots)$.
Applying the Schur-complement formula for the upper-left block gives
\begin{equation}\begin{split}
&\langle n+1| \left(pI_{n+1:\infty}+i\mathcal{J}_{n+1:\infty}\right)^{-1}|n+1\rangle
\\
=& \big[p+i\omega_{n+1}
\\
&+\kappa_{n+1}^2\langle n+2|\left(pI_{n+2:\infty}+i\mathcal{J}_{n+2:\infty}\right)^{-1}|n+2\rangle \big]^{-1}.
\end{split}\end{equation}
Using the self-energy definition at the next port,
\begin{equation}
\begin{split}
&\Sigma_{n+2\to n+1}(p)
\\
&=\kappa_{n+1}^2\langle n+2|\left(pI_{n+2:\infty}+i\mathcal{J}_{n+2:\infty}\right)^{-1}|n+2\rangle,
\end{split}
\end{equation}
the $n+1\to n$ self-energy is given by the recurrence relation
\begin{equation}\label{eq:BSE recurrence right}
\Sigma_{n+1\to n}(p) =\kappa_n^2\left[p+i\omega_{n+1}+\Sigma_{n+2\to n+1}(p)\right]^{-1}.
\end{equation}

The diagonal resolvent matrix elements appearing in the definitions of the two self-energies are the surface Green functions of the structures connected to the corresponding ports.
For $n\geq1$, the local Green function is
\begin{equation}
G_{nn}(p)=\frac{1}{p+i\omega_n+\Sigma_{n-1\to n}(p)+\Sigma_{n+1\to n}(p)}.
\end{equation}
At the qubit endpoint there is only one boundary port, so
\begin{equation}
G_{00}(p)=\frac{1}{p+i\epsilon+\Sigma_{1\to0}(p)}.
\end{equation}
Thus, the two directional self-energies enter additively into the inverse local Green function.

\section{Spectral density, self-energy, and chain representation}
\label{app:star_chain_bse}

In this appendix, we show that the star and chain representations of the environment yield the same self-energy at the qubit port and introduce the construction of the chain coefficients.

\subsection{Star and chain representations}

Let $D_{\mu\nu}=\Omega_\mu\delta_{\mu\nu}$ and $\mathbf h=(h_1,h_2,\ldots)^T$ denote the star-bath frequency matrix and coupling vector, with the convention of Eq.~\eqref{eq:B_star_integral}.
The exact chain representation is a unitary change of bath basis~\cite{woods2014mappings,chin2010exact}.
On the closed bath subspace generated by $\mathbf h,D\mathbf h,D^2\mathbf h,\ldots$, the transformation $U$ gives
\begin{equation}\label{eq:B_unitary_mapping}
\mathcal J_{1:\infty}=UDU^\dagger,\qquad U\mathbf h=g|1\rangle,
\end{equation}
where $D$ is restricted to this subspace, $|1\rangle$ is the first chain-site basis vector, and $g^2=\mathbf h^\dagger\mathbf h$.
Modes outside this subspace do not contribute to the qubit-port self-energy.
The corresponding resolvents satisfy
\begin{equation}\label{eq:B_resolvent_unitary}
(pI+i\mathcal J_{1:\infty})^{-1}=U(pI+iD)^{-1}U^\dagger.
\end{equation}
Thus the self-energy in Eq.~\eqref{eq:B_star_integral} is unchanged by the mapping,
\begin{equation}\label{eq:B_star_chain_equivalence}
\begin{split}
\Sigma_\star(p)&=\mathbf h^\dagger(pI+iD)^{-1}\mathbf h
\\
&=g^2\langle1|(pI+i\mathcal J_{1:\infty})^{-1}|1\rangle
\\
&=\Sigma_{1\to0}^{\mathrm{exact}}(p).
\end{split}
\end{equation}
This identity justifies using the continuum response as the reference for the terminated chains.
For $K$ discrete bath modes, the chain has at most $K$ sites; for the continuous bath considered here, it is semi-infinite.

\subsection{Chain coefficients and spectral normalization}

The chain coefficients are determined by the spectral density through the orthogonal-polynomial chain mapping~\cite{chin2010exact,woods2014mappings}.
We introduce the normalized coupling-weighted measure
\begin{equation}\label{eq:B_measure}
d\mu(\nu)=\frac{J(\nu)}{\pi g^2}d\nu,\qquad g^2=\frac{1}{\pi}\int_{\omega_L}^{\omega_R}J(\nu)d\nu,
\end{equation}
so that $\int_{\omega_L}^{\omega_R}d\mu(\nu)=1$.
Let $P_n(\nu)$ be the polynomials of degree $n$, with positive leading coefficients, that are orthonormal with respect to $d\mu$, so that $P_0(\nu)=1$.
They generate the Jacobi matrix of the semi-infinite chain, whose coefficients are
\begin{equation}\label{eq:B_chain_coefficients}
\begin{split}
\omega_n&=\int_{\omega_L}^{\omega_R}\nu P_{n-1}(\nu)^2d\mu(\nu),
\\
\kappa_n&=\int_{\omega_L}^{\omega_R}\nu P_{n-1}(\nu)P_n(\nu)d\mu(\nu).
\end{split}
\end{equation}
We compute the chain coefficients numerically by Lanczos tridiagonalization, with full re-orthogonalization, of a 4000-node Gauss--Legendre discretization of $d\mu$.
For the benchmark continuum bath, the spectral density is positive almost everywhere on $[\omega_L,\omega_R]=[0,1]$, and the exact chain coefficients approach the asymptotic values~\cite{woods2014mappings}
\begin{equation}
\omega_\infty=\frac{\omega_L+\omega_R}{2}=\frac{1}{2},\qquad \kappa_\infty=\frac{\omega_R-\omega_L}{4}=\frac{1}{4}.
\end{equation}

For the structured spectral density in Eq.~\eqref{chain_spectral_density}, the normalization is
\begin{equation}\label{eq:B_spectral_normalization}
\mathcal N=2\int_0^1\nu\left[1+A e^{-(\nu-\omega_0)^2/(2\sigma^2)}\right]d\nu.
\end{equation}
This gives $g^2=\pi^{-1}\int_0^1J(\nu)d\nu=\alpha$, as used in Sec.~\ref{Numerical_demonstration}.

\bibliography{references}

\begin{thebibliography}{72}%
\makeatletter
\providecommand \@ifxundefined [1]{%
 \@ifx{#1\undefined}
}%
\providecommand \@ifnum [1]{%
 \ifnum #1\expandafter \@firstoftwo
 \else \expandafter \@secondoftwo
 \fi
}%
\providecommand \@ifx [1]{%
 \ifx #1\expandafter \@firstoftwo
 \else \expandafter \@secondoftwo
 \fi
}%
\providecommand \natexlab [1]{#1}%
\providecommand \enquote  [1]{``#1''}%
\providecommand \bibnamefont  [1]{#1}%
\providecommand \bibfnamefont [1]{#1}%
\providecommand \citenamefont [1]{#1}%
\providecommand \href@noop [0]{\@secondoftwo}%
\providecommand \href [0]{\begingroup \@sanitize@url \@href}%
\providecommand \@href[1]{\@@startlink{#1}\@@href}%
\providecommand \@@href[1]{\endgroup#1\@@endlink}%
\providecommand \@sanitize@url [0]{\catcode `\\12\catcode `\$12\catcode
  `\&12\catcode `\#12\catcode `\^12\catcode `\_12\catcode `\%12\relax}%
\providecommand \@@startlink[1]{}%
\providecommand \@@endlink[0]{}%
\providecommand \url  [0]{\begingroup\@sanitize@url \@url }%
\providecommand \@url [1]{\endgroup\@href {#1}{\urlprefix }}%
\providecommand \urlprefix  [0]{URL }%
\providecommand \Eprint [0]{\href }%
\providecommand \doibase [0]{https://doi.org/}%
\providecommand \selectlanguage [0]{\@gobble}%
\providecommand \bibinfo  [0]{\@secondoftwo}%
\providecommand \bibfield  [0]{\@secondoftwo}%
\providecommand \translation [1]{[#1]}%
\providecommand \BibitemOpen [0]{}%
\providecommand \bibitemStop [0]{}%
\providecommand \bibitemNoStop [0]{.\EOS\space}%
\providecommand \EOS [0]{\spacefactor3000\relax}%
\providecommand \BibitemShut  [1]{\csname bibitem#1\endcsname}%
\let\auto@bib@innerbib\@empty
\bibitem [{\citenamefont {Breuer}\ and\ \citenamefont
  {Petruccione}(2002)}]{breuer2002theory}%
  \BibitemOpen
  \bibfield  {author} {\bibinfo {author} {\bibfnamefont {H.-P.}\ \bibnamefont
  {Breuer}}\ and\ \bibinfo {author} {\bibfnamefont {F.}~\bibnamefont
  {Petruccione}},\ }\href@noop {} {\emph {\bibinfo {title} {The theory of open
  quantum systems}}}\ (\bibinfo  {publisher} {Oxford University Press, USA},\
  \bibinfo {year} {2002})\BibitemShut {NoStop}%
\bibitem [{\citenamefont {Gardiner}\ and\ \citenamefont
  {Zoller}(2004)}]{gardiner2004quantum}%
  \BibitemOpen
  \bibfield  {author} {\bibinfo {author} {\bibfnamefont {C.}~\bibnamefont
  {Gardiner}}\ and\ \bibinfo {author} {\bibfnamefont {P.}~\bibnamefont
  {Zoller}},\ }\href@noop {} {\emph {\bibinfo {title} {Quantum noise: a
  handbook of {Markovian} and non-{Markovian} quantum stochastic methods with
  applications to quantum optics}}}\ (\bibinfo  {publisher} {Springer Science
  \& Business Media},\ \bibinfo {year} {2004})\BibitemShut {NoStop}%
\bibitem [{\citenamefont {Nakajima}(1958)}]{nakajima1958quantum}%
  \BibitemOpen
  \bibfield  {author} {\bibinfo {author} {\bibfnamefont {S.}~\bibnamefont
  {Nakajima}},\ }\bibfield  {title} {\bibinfo {title} {On quantum theory of
  transport phenomena: Steady diffusion},\ }\href@noop {} {\bibfield  {journal}
  {\bibinfo  {journal} {Prog. Theor. Phys.}\ }\textbf {\bibinfo {volume}
  {20}},\ \bibinfo {pages} {948} (\bibinfo {year} {1958})}\BibitemShut
  {NoStop}%
\bibitem [{\citenamefont {Zwanzig}(1960)}]{zwanzig1960ensemble}%
  \BibitemOpen
  \bibfield  {author} {\bibinfo {author} {\bibfnamefont {R.}~\bibnamefont
  {Zwanzig}},\ }\bibfield  {title} {\bibinfo {title} {Ensemble method in the
  theory of irreversibility},\ }\href@noop {} {\bibfield  {journal} {\bibinfo
  {journal} {J. Chem. Phys.}\ }\textbf {\bibinfo {volume} {33}},\ \bibinfo
  {pages} {1338} (\bibinfo {year} {1960})}\BibitemShut {NoStop}%
\bibitem [{\citenamefont {Feynman}\ and\ \citenamefont
  {Vernon}(1963)}]{FEYNMAN1963118}%
  \BibitemOpen
  \bibfield  {author} {\bibinfo {author} {\bibfnamefont {R.~P.}\ \bibnamefont
  {Feynman}}\ and\ \bibinfo {author} {\bibfnamefont {F.~L.}\ \bibnamefont
  {Vernon}},\ }\bibfield  {title} {\bibinfo {title} {The theory of a general
  quantum system interacting with a linear dissipative system},\ }\href
  {https://doi.org/10.1016/0003-4916(63)90068-X} {\bibfield  {journal}
  {\bibinfo  {journal} {Ann. Phys.}\ }\textbf {\bibinfo {volume} {24}},\
  \bibinfo {pages} {118} (\bibinfo {year} {1963})}\BibitemShut {NoStop}%
\bibitem [{\citenamefont {Gorini}\ \emph {et~al.}(1976)\citenamefont {Gorini},
  \citenamefont {Kossakowski},\ and\ \citenamefont
  {Sudarshan}}]{gorini1976completely}%
  \BibitemOpen
  \bibfield  {author} {\bibinfo {author} {\bibfnamefont {V.}~\bibnamefont
  {Gorini}}, \bibinfo {author} {\bibfnamefont {A.}~\bibnamefont
  {Kossakowski}},\ and\ \bibinfo {author} {\bibfnamefont {E.~C.~G.}\
  \bibnamefont {Sudarshan}},\ }\bibfield  {title} {\bibinfo {title} {Completely
  positive dynamical semigroups of {N}-level systems},\ }\href@noop {}
  {\bibfield  {journal} {\bibinfo  {journal} {J. Math. Phys.}\ }\textbf
  {\bibinfo {volume} {17}},\ \bibinfo {pages} {821} (\bibinfo {year}
  {1976})}\BibitemShut {NoStop}%
\bibitem [{\citenamefont {Lindblad}(1976)}]{lindblad1976generators}%
  \BibitemOpen
  \bibfield  {author} {\bibinfo {author} {\bibfnamefont {G.}~\bibnamefont
  {Lindblad}},\ }\bibfield  {title} {\bibinfo {title} {On the generators of
  quantum dynamical semigroups},\ }\href@noop {} {\bibfield  {journal}
  {\bibinfo  {journal} {Commun. Math. Phys.}\ }\textbf {\bibinfo {volume}
  {48}},\ \bibinfo {pages} {119} (\bibinfo {year} {1976})}\BibitemShut
  {NoStop}%
\bibitem [{\citenamefont {S{\'a}nchez-Barquilla}\ and\ \citenamefont
  {Feist}(2021)}]{sanchez2021accurate}%
  \BibitemOpen
  \bibfield  {author} {\bibinfo {author} {\bibfnamefont {M.}~\bibnamefont
  {S{\'a}nchez-Barquilla}}\ and\ \bibinfo {author} {\bibfnamefont
  {J.}~\bibnamefont {Feist}},\ }\bibfield  {title} {\bibinfo {title} {Accurate
  truncations of chain mapping models for open quantum systems},\ }\href@noop
  {} {\bibfield  {journal} {\bibinfo  {journal} {Nanomaterials}\ }\textbf
  {\bibinfo {volume} {11}},\ \bibinfo {pages} {2104} (\bibinfo {year}
  {2021})}\BibitemShut {NoStop}%
\bibitem [{\citenamefont {Rivas}\ \emph {et~al.}(2014)\citenamefont {Rivas},
  \citenamefont {Huelga},\ and\ \citenamefont {Plenio}}]{rivas2014quantum}%
  \BibitemOpen
  \bibfield  {author} {\bibinfo {author} {\bibfnamefont {{\'A}.}~\bibnamefont
  {Rivas}}, \bibinfo {author} {\bibfnamefont {S.~F.}\ \bibnamefont {Huelga}},\
  and\ \bibinfo {author} {\bibfnamefont {M.~B.}\ \bibnamefont {Plenio}},\
  }\bibfield  {title} {\bibinfo {title} {Quantum non-{Markovianity}:
  characterization, quantification and detection},\ }\href@noop {} {\bibfield
  {journal} {\bibinfo  {journal} {Rep. Prog. Phys.}\ }\textbf {\bibinfo
  {volume} {77}},\ \bibinfo {pages} {094001} (\bibinfo {year}
  {2014})}\BibitemShut {NoStop}%
\bibitem [{\citenamefont {Breuer}\ \emph {et~al.}(2016)\citenamefont {Breuer},
  \citenamefont {Laine}, \citenamefont {Piilo},\ and\ \citenamefont
  {Vacchini}}]{breuer2016colloquium}%
  \BibitemOpen
  \bibfield  {author} {\bibinfo {author} {\bibfnamefont {H.-P.}\ \bibnamefont
  {Breuer}}, \bibinfo {author} {\bibfnamefont {E.-M.}\ \bibnamefont {Laine}},
  \bibinfo {author} {\bibfnamefont {J.}~\bibnamefont {Piilo}},\ and\ \bibinfo
  {author} {\bibfnamefont {B.}~\bibnamefont {Vacchini}},\ }\bibfield  {title}
  {\bibinfo {title} {Colloquium: Non-{Markovian} dynamics in open quantum
  systems},\ }\href@noop {} {\bibfield  {journal} {\bibinfo  {journal} {Rev.
  Mod. Phys.}\ }\textbf {\bibinfo {volume} {88}},\ \bibinfo {pages} {021002}
  (\bibinfo {year} {2016})}\BibitemShut {NoStop}%
\bibitem [{\citenamefont {de~Vega}\ and\ \citenamefont
  {Alonso}(2017)}]{de2017dynamics}%
  \BibitemOpen
  \bibfield  {author} {\bibinfo {author} {\bibfnamefont {I.}~\bibnamefont
  {de~Vega}}\ and\ \bibinfo {author} {\bibfnamefont {D.}~\bibnamefont
  {Alonso}},\ }\bibfield  {title} {\bibinfo {title} {Dynamics of
  non-{Markovian} open quantum systems},\ }\href@noop {} {\bibfield  {journal}
  {\bibinfo  {journal} {Rev. Mod. Phys.}\ }\textbf {\bibinfo {volume} {89}},\
  \bibinfo {pages} {015001} (\bibinfo {year} {2017})}\BibitemShut {NoStop}%
\bibitem [{\citenamefont {Zhao}\ \emph {et~al.}(2011)\citenamefont {Zhao},
  \citenamefont {Jing}, \citenamefont {Corn},\ and\ \citenamefont
  {Yu}}]{zhao2011dynamics}%
  \BibitemOpen
  \bibfield  {author} {\bibinfo {author} {\bibfnamefont {X.}~\bibnamefont
  {Zhao}}, \bibinfo {author} {\bibfnamefont {J.}~\bibnamefont {Jing}}, \bibinfo
  {author} {\bibfnamefont {B.}~\bibnamefont {Corn}},\ and\ \bibinfo {author}
  {\bibfnamefont {T.}~\bibnamefont {Yu}},\ }\bibfield  {title} {\bibinfo
  {title} {Dynamics of interacting qubits coupled to a common bath:
  Non-{Markovian} quantum-state-diffusion approach},\ }\href@noop {} {\bibfield
   {journal} {\bibinfo  {journal} {Phys. Rev. A}\ }\textbf {\bibinfo {volume}
  {84}},\ \bibinfo {pages} {032101} (\bibinfo {year} {2011})}\BibitemShut
  {NoStop}%
\bibitem [{\citenamefont {Mu}\ \emph {et~al.}(2016)\citenamefont {Mu},
  \citenamefont {Zhao},\ and\ \citenamefont {Yu}}]{mu2016memory}%
  \BibitemOpen
  \bibfield  {author} {\bibinfo {author} {\bibfnamefont {Q.}~\bibnamefont
  {Mu}}, \bibinfo {author} {\bibfnamefont {X.}~\bibnamefont {Zhao}},\ and\
  \bibinfo {author} {\bibfnamefont {T.}~\bibnamefont {Yu}},\ }\bibfield
  {title} {\bibinfo {title} {{Memory-effect-induced macroscopic-microscopic
  entanglement}},\ }\href@noop {} {\bibfield  {journal} {\bibinfo  {journal}
  {Phys. Rev. A}\ }\textbf {\bibinfo {volume} {94}},\ \bibinfo {pages} {012334}
  (\bibinfo {year} {2016})}\BibitemShut {NoStop}%
\bibitem [{\citenamefont {Luo}\ \emph {et~al.}(2018)\citenamefont {Luo},
  \citenamefont {You}, \citenamefont {Lin}, \citenamefont {Wu},\ and\
  \citenamefont {Yu}}]{luo2018memory}%
  \BibitemOpen
  \bibfield  {author} {\bibinfo {author} {\bibfnamefont {D.-W.}\ \bibnamefont
  {Luo}}, \bibinfo {author} {\bibfnamefont {J.~Q.}\ \bibnamefont {You}},
  \bibinfo {author} {\bibfnamefont {H.-Q.}\ \bibnamefont {Lin}}, \bibinfo
  {author} {\bibfnamefont {L.-A.}\ \bibnamefont {Wu}},\ and\ \bibinfo {author}
  {\bibfnamefont {T.}~\bibnamefont {Yu}},\ }\bibfield  {title} {\bibinfo
  {title} {Memory-induced geometric phase in non-{Markovian} open systems},\
  }\href@noop {} {\bibfield  {journal} {\bibinfo  {journal} {Phys. Rev. A}\
  }\textbf {\bibinfo {volume} {98}},\ \bibinfo {pages} {052117} (\bibinfo
  {year} {2018})}\BibitemShut {NoStop}%
\bibitem [{\citenamefont {Xiong}\ \emph {et~al.}(2015)\citenamefont {Xiong},
  \citenamefont {Lo}, \citenamefont {Zhang}, \citenamefont {Feng},\ and\
  \citenamefont {Nori}}]{xiong2015non}%
  \BibitemOpen
  \bibfield  {author} {\bibinfo {author} {\bibfnamefont {H.-N.}\ \bibnamefont
  {Xiong}}, \bibinfo {author} {\bibfnamefont {P.-Y.}\ \bibnamefont {Lo}},
  \bibinfo {author} {\bibfnamefont {W.-M.}\ \bibnamefont {Zhang}}, \bibinfo
  {author} {\bibfnamefont {D.~H.}\ \bibnamefont {Feng}},\ and\ \bibinfo
  {author} {\bibfnamefont {F.}~\bibnamefont {Nori}},\ }\bibfield  {title}
  {\bibinfo {title} {Non-{Markovian} complexity in the quantum-to-classical
  transition},\ }\href@noop {} {\bibfield  {journal} {\bibinfo  {journal} {Sci.
  Rep.}\ }\textbf {\bibinfo {volume} {5}},\ \bibinfo {pages} {13353} (\bibinfo
  {year} {2015})}\BibitemShut {NoStop}%
\bibitem [{\citenamefont {Yu}\ \emph {et~al.}(1999)\citenamefont {Yu},
  \citenamefont {Di{\'o}si}, \citenamefont {Gisin},\ and\ \citenamefont
  {Strunz}}]{yu1999non}%
  \BibitemOpen
  \bibfield  {author} {\bibinfo {author} {\bibfnamefont {T.}~\bibnamefont
  {Yu}}, \bibinfo {author} {\bibfnamefont {L.}~\bibnamefont {Di{\'o}si}},
  \bibinfo {author} {\bibfnamefont {N.}~\bibnamefont {Gisin}},\ and\ \bibinfo
  {author} {\bibfnamefont {W.~T.}\ \bibnamefont {Strunz}},\ }\bibfield  {title}
  {\bibinfo {title} {Non-{Markovian} quantum-state diffusion: Perturbation
  approach},\ }\href@noop {} {\bibfield  {journal} {\bibinfo  {journal} {Phys.
  Rev. A}\ }\textbf {\bibinfo {volume} {60}},\ \bibinfo {pages} {91} (\bibinfo
  {year} {1999})}\BibitemShut {NoStop}%
\bibitem [{\citenamefont {Strunz}\ and\ \citenamefont
  {Yu}(2004)}]{strunz2004convolutionless}%
  \BibitemOpen
  \bibfield  {author} {\bibinfo {author} {\bibfnamefont {W.~T.}\ \bibnamefont
  {Strunz}}\ and\ \bibinfo {author} {\bibfnamefont {T.}~\bibnamefont {Yu}},\
  }\bibfield  {title} {\bibinfo {title} {Convolutionless non-{Markovian} master
  equations and quantum trajectories: {Brownian} motion},\ }\href@noop {}
  {\bibfield  {journal} {\bibinfo  {journal} {Phys. Rev. A}\ }\textbf {\bibinfo
  {volume} {69}},\ \bibinfo {pages} {052115} (\bibinfo {year}
  {2004})}\BibitemShut {NoStop}%
\bibitem [{\citenamefont {Chen}\ \emph {et~al.}(2014)\citenamefont {Chen},
  \citenamefont {You},\ and\ \citenamefont {Yu}}]{chen2014exact}%
  \BibitemOpen
  \bibfield  {author} {\bibinfo {author} {\bibfnamefont {Y.}~\bibnamefont
  {Chen}}, \bibinfo {author} {\bibfnamefont {J.~Q.}\ \bibnamefont {You}},\ and\
  \bibinfo {author} {\bibfnamefont {T.}~\bibnamefont {Yu}},\ }\bibfield
  {title} {\bibinfo {title} {Exact non-{M}arkovian master equations for
  multiple qubit systems: Quantum-trajectory approach},\ }\href
  {https://doi.org/10.1103/PhysRevA.90.052104} {\bibfield  {journal} {\bibinfo
  {journal} {Phys. Rev. A}\ }\textbf {\bibinfo {volume} {90}},\ \bibinfo
  {pages} {052104} (\bibinfo {year} {2014})}\BibitemShut {NoStop}%
\bibitem [{\citenamefont {Xu}\ \emph {et~al.}(2014)\citenamefont {Xu},
  \citenamefont {Zhao}, \citenamefont {Jing}, \citenamefont {Wu},\ and\
  \citenamefont {Yu}}]{xu2014perturbation}%
  \BibitemOpen
  \bibfield  {author} {\bibinfo {author} {\bibfnamefont {J.}~\bibnamefont
  {Xu}}, \bibinfo {author} {\bibfnamefont {X.}~\bibnamefont {Zhao}}, \bibinfo
  {author} {\bibfnamefont {J.}~\bibnamefont {Jing}}, \bibinfo {author}
  {\bibfnamefont {L.~A.}\ \bibnamefont {Wu}},\ and\ \bibinfo {author}
  {\bibfnamefont {T.}~\bibnamefont {Yu}},\ }\bibfield  {title} {\bibinfo
  {title} {Perturbation methods for the non-{Markovian} quantum state diffusion
  equation},\ }\href@noop {} {\bibfield  {journal} {\bibinfo  {journal} {J.
  Phys. A: Math. Theor.}\ }\textbf {\bibinfo {volume} {47}},\ \bibinfo {pages}
  {435301} (\bibinfo {year} {2014})}\BibitemShut {NoStop}%
\bibitem [{\citenamefont {Hu}\ \emph {et~al.}(1992)\citenamefont {Hu},
  \citenamefont {Paz},\ and\ \citenamefont {Zhang}}]{hu1992quantum}%
  \BibitemOpen
  \bibfield  {author} {\bibinfo {author} {\bibfnamefont {B.~L.}\ \bibnamefont
  {Hu}}, \bibinfo {author} {\bibfnamefont {J.~P.}\ \bibnamefont {Paz}},\ and\
  \bibinfo {author} {\bibfnamefont {Y.}~\bibnamefont {Zhang}},\ }\bibfield
  {title} {\bibinfo {title} {Quantum {Brownian} motion in a general
  environment: Exact master equation with nonlocal dissipation and colored
  noise},\ }\href@noop {} {\bibfield  {journal} {\bibinfo  {journal} {Phys.
  Rev. D}\ }\textbf {\bibinfo {volume} {45}},\ \bibinfo {pages} {2843}
  (\bibinfo {year} {1992})}\BibitemShut {NoStop}%
\bibitem [{\citenamefont {An}\ and\ \citenamefont {Zhang}(2007)}]{an2007non}%
  \BibitemOpen
  \bibfield  {author} {\bibinfo {author} {\bibfnamefont {J.-H.}\ \bibnamefont
  {An}}\ and\ \bibinfo {author} {\bibfnamefont {W.-M.}\ \bibnamefont {Zhang}},\
  }\bibfield  {title} {\bibinfo {title} {Non-{Markovian} entanglement dynamics
  of noisy continuous-variable quantum channels},\ }\href@noop {} {\bibfield
  {journal} {\bibinfo  {journal} {Phys. Rev. A}\ }\textbf {\bibinfo {volume}
  {76}},\ \bibinfo {pages} {042127} (\bibinfo {year} {2007})}\BibitemShut
  {NoStop}%
\bibitem [{\citenamefont {Zhang}\ \emph {et~al.}(2012)\citenamefont {Zhang},
  \citenamefont {Lo}, \citenamefont {Xiong}, \citenamefont {Tu},\ and\
  \citenamefont {Nori}}]{zhang2012general}%
  \BibitemOpen
  \bibfield  {author} {\bibinfo {author} {\bibfnamefont {W.-M.}\ \bibnamefont
  {Zhang}}, \bibinfo {author} {\bibfnamefont {P.-Y.}\ \bibnamefont {Lo}},
  \bibinfo {author} {\bibfnamefont {H.-N.}\ \bibnamefont {Xiong}}, \bibinfo
  {author} {\bibfnamefont {M.~W.-Y.}\ \bibnamefont {Tu}},\ and\ \bibinfo
  {author} {\bibfnamefont {F.}~\bibnamefont {Nori}},\ }\bibfield  {title}
  {\bibinfo {title} {General non-{Markovian} dynamics of open quantum
  systems},\ }\href@noop {} {\bibfield  {journal} {\bibinfo  {journal} {Phys.
  Rev. Lett.}\ }\textbf {\bibinfo {volume} {109}},\ \bibinfo {pages} {170402}
  (\bibinfo {year} {2012})}\BibitemShut {NoStop}%
\bibitem [{\citenamefont {Zhang}(2019)}]{zhang2019exact}%
  \BibitemOpen
  \bibfield  {author} {\bibinfo {author} {\bibfnamefont {W.-M.}\ \bibnamefont
  {Zhang}},\ }\bibfield  {title} {\bibinfo {title} {Exact master equation and
  general non-{Markovian} dynamics in open quantum systems},\ }\href@noop {}
  {\bibfield  {journal} {\bibinfo  {journal} {Eur. Phys. J. Spec. Top.}\
  }\textbf {\bibinfo {volume} {227}},\ \bibinfo {pages} {1849} (\bibinfo {year}
  {2019})}\BibitemShut {NoStop}%
\bibitem [{\citenamefont {Xu}\ \emph {et~al.}(2026)\citenamefont {Xu},
  \citenamefont {Vadimov}, \citenamefont {Stockburger},\ and\ \citenamefont
  {Ankerhold}}]{xu2026colloquium}%
  \BibitemOpen
  \bibfield  {author} {\bibinfo {author} {\bibfnamefont {M.}~\bibnamefont
  {Xu}}, \bibinfo {author} {\bibfnamefont {V.}~\bibnamefont {Vadimov}},
  \bibinfo {author} {\bibfnamefont {J.}~\bibnamefont {Stockburger}},\ and\
  \bibinfo {author} {\bibfnamefont {J.}~\bibnamefont {Ankerhold}},\ }\bibfield
  {title} {\bibinfo {title} {Colloquium: Simulating non-{M}arkovian dynamics in
  open quantum systems},\ }\href@noop {} {\bibfield  {journal} {\bibinfo
  {journal} {Reviews of Modern Physics}\ }\textbf {\bibinfo {volume} {98}},\
  \bibinfo {pages} {021002} (\bibinfo {year} {2026})}\BibitemShut {NoStop}%
\bibitem [{\citenamefont {Ma}\ \emph {et~al.}(2014)\citenamefont {Ma},
  \citenamefont {Chen}, \citenamefont {Chen}, \citenamefont {Hedemann},\ and\
  \citenamefont {Yu}}]{ma2014crossover}%
  \BibitemOpen
  \bibfield  {author} {\bibinfo {author} {\bibfnamefont {T.}~\bibnamefont
  {Ma}}, \bibinfo {author} {\bibfnamefont {Y.}~\bibnamefont {Chen}}, \bibinfo
  {author} {\bibfnamefont {T.}~\bibnamefont {Chen}}, \bibinfo {author}
  {\bibfnamefont {S.~R.}\ \bibnamefont {Hedemann}},\ and\ \bibinfo {author}
  {\bibfnamefont {T.}~\bibnamefont {Yu}},\ }\bibfield  {title} {\bibinfo
  {title} {Crossover between non-{Markovian} and {Markovian} dynamics induced
  by a hierarchical environment},\ }\href@noop {} {\bibfield  {journal}
  {\bibinfo  {journal} {Phys. Rev. A}\ }\textbf {\bibinfo {volume} {90}},\
  \bibinfo {pages} {042108} (\bibinfo {year} {2014})}\BibitemShut {NoStop}%
\bibitem [{\citenamefont {Di{\'o}si}\ \emph {et~al.}(1998)\citenamefont
  {Di{\'o}si}, \citenamefont {Gisin},\ and\ \citenamefont
  {Strunz}}]{diosi1998non}%
  \BibitemOpen
  \bibfield  {author} {\bibinfo {author} {\bibfnamefont {L.}~\bibnamefont
  {Di{\'o}si}}, \bibinfo {author} {\bibfnamefont {N.}~\bibnamefont {Gisin}},\
  and\ \bibinfo {author} {\bibfnamefont {W.~T.}\ \bibnamefont {Strunz}},\
  }\bibfield  {title} {\bibinfo {title} {{Non-Markovian quantum state
  diffusion}},\ }\href@noop {} {\bibfield  {journal} {\bibinfo  {journal}
  {Phys. Rev. A}\ }\textbf {\bibinfo {volume} {58}},\ \bibinfo {pages} {1699}
  (\bibinfo {year} {1998})}\BibitemShut {NoStop}%
\bibitem [{\citenamefont {Garg}\ \emph {et~al.}(1985)\citenamefont {Garg},
  \citenamefont {Onuchic},\ and\ \citenamefont {Ambegaokar}}]{garg1985effect}%
  \BibitemOpen
  \bibfield  {author} {\bibinfo {author} {\bibfnamefont {A.}~\bibnamefont
  {Garg}}, \bibinfo {author} {\bibfnamefont {J.~N.}\ \bibnamefont {Onuchic}},\
  and\ \bibinfo {author} {\bibfnamefont {V.}~\bibnamefont {Ambegaokar}},\
  }\bibfield  {title} {\bibinfo {title} {Effect of friction on electron
  transfer in biomolecules},\ }\href@noop {} {\bibfield  {journal} {\bibinfo
  {journal} {The Journal of Chemical Physics}\ }\textbf {\bibinfo {volume}
  {83}},\ \bibinfo {pages} {4491} (\bibinfo {year} {1985})}\BibitemShut
  {NoStop}%
\bibitem [{\citenamefont {Iles-Smith}\ \emph {et~al.}(2014)\citenamefont
  {Iles-Smith}, \citenamefont {Lambert},\ and\ \citenamefont
  {Nazir}}]{iles2014environmental}%
  \BibitemOpen
  \bibfield  {author} {\bibinfo {author} {\bibfnamefont {J.}~\bibnamefont
  {Iles-Smith}}, \bibinfo {author} {\bibfnamefont {N.}~\bibnamefont
  {Lambert}},\ and\ \bibinfo {author} {\bibfnamefont {A.}~\bibnamefont
  {Nazir}},\ }\bibfield  {title} {\bibinfo {title} {Environmental dynamics,
  correlations, and the emergence of noncanonical equilibrium states in open
  quantum systems},\ }\href@noop {} {\bibfield  {journal} {\bibinfo  {journal}
  {Phys. Rev. A}\ }\textbf {\bibinfo {volume} {90}},\ \bibinfo {pages} {032114}
  (\bibinfo {year} {2014})}\BibitemShut {NoStop}%
\bibitem [{\citenamefont {Strasberg}\ \emph {et~al.}(2016)\citenamefont
  {Strasberg}, \citenamefont {Schaller}, \citenamefont {Lambert},\ and\
  \citenamefont {Brandes}}]{strasberg2016nonequilibrium}%
  \BibitemOpen
  \bibfield  {author} {\bibinfo {author} {\bibfnamefont {P.}~\bibnamefont
  {Strasberg}}, \bibinfo {author} {\bibfnamefont {G.}~\bibnamefont {Schaller}},
  \bibinfo {author} {\bibfnamefont {N.}~\bibnamefont {Lambert}},\ and\ \bibinfo
  {author} {\bibfnamefont {T.}~\bibnamefont {Brandes}},\ }\bibfield  {title}
  {\bibinfo {title} {Nonequilibrium thermodynamics in the strong coupling and
  non-{M}arkovian regime based on a reaction coordinate mapping},\ }\href@noop
  {} {\bibfield  {journal} {\bibinfo  {journal} {New Journal of Physics}\
  }\textbf {\bibinfo {volume} {18}},\ \bibinfo {pages} {073007} (\bibinfo
  {year} {2016})}\BibitemShut {NoStop}%
\bibitem [{\citenamefont {Nazir}\ and\ \citenamefont
  {Schaller}(2019)}]{nazir2019reaction}%
  \BibitemOpen
  \bibfield  {author} {\bibinfo {author} {\bibfnamefont {A.}~\bibnamefont
  {Nazir}}\ and\ \bibinfo {author} {\bibfnamefont {G.}~\bibnamefont
  {Schaller}},\ }\bibfield  {title} {\bibinfo {title} {The reaction coordinate
  mapping in quantum thermodynamics},\ }in\ \href@noop {} {\emph {\bibinfo
  {booktitle} {Thermodynamics in the Quantum Regime: Fundamental Aspects and
  New Directions}}}\ (\bibinfo  {publisher} {Springer},\ \bibinfo {year}
  {2019})\ pp.\ \bibinfo {pages} {551--577}\BibitemShut {NoStop}%
\bibitem [{\citenamefont {Chin}\ \emph {et~al.}(2010)\citenamefont {Chin},
  \citenamefont {Rivas}, \citenamefont {Huelga},\ and\ \citenamefont
  {Plenio}}]{chin2010exact}%
  \BibitemOpen
  \bibfield  {author} {\bibinfo {author} {\bibfnamefont {A.~W.}\ \bibnamefont
  {Chin}}, \bibinfo {author} {\bibfnamefont {{\'A}.}~\bibnamefont {Rivas}},
  \bibinfo {author} {\bibfnamefont {S.~F.}\ \bibnamefont {Huelga}},\ and\
  \bibinfo {author} {\bibfnamefont {M.~B.}\ \bibnamefont {Plenio}},\ }\bibfield
   {title} {\bibinfo {title} {Exact mapping between system-reservoir quantum
  models and semi-infinite discrete chains using orthogonal polynomials},\
  }\href@noop {} {\bibfield  {journal} {\bibinfo  {journal} {J. Math. Phys.}\
  }\textbf {\bibinfo {volume} {51}},\ \bibinfo {pages} {092109} (\bibinfo
  {year} {2010})}\BibitemShut {NoStop}%
\bibitem [{\citenamefont {Martinazzo}\ \emph
  {et~al.}(2011{\natexlab{a}})\citenamefont {Martinazzo}, \citenamefont
  {Vacchini}, \citenamefont {Hughes},\ and\ \citenamefont
  {Burghardt}}]{martinazzo2011communication}%
  \BibitemOpen
  \bibfield  {author} {\bibinfo {author} {\bibfnamefont {R.}~\bibnamefont
  {Martinazzo}}, \bibinfo {author} {\bibfnamefont {B.}~\bibnamefont
  {Vacchini}}, \bibinfo {author} {\bibfnamefont {K.~H.}\ \bibnamefont
  {Hughes}},\ and\ \bibinfo {author} {\bibfnamefont {I.}~\bibnamefont
  {Burghardt}},\ }\bibfield  {title} {\bibinfo {title} {Communication:
  Universal {Markovian} reduction of {Brownian} particle dynamics},\
  }\href@noop {} {\bibfield  {journal} {\bibinfo  {journal} {J. Chem. Phys.}\
  }\textbf {\bibinfo {volume} {134}},\ \bibinfo {pages} {011101} (\bibinfo
  {year} {2011}{\natexlab{a}})}\BibitemShut {NoStop}%
\bibitem [{\citenamefont {Woods}\ \emph {et~al.}(2014)\citenamefont {Woods},
  \citenamefont {Groux}, \citenamefont {Chin}, \citenamefont {Huelga},\ and\
  \citenamefont {Plenio}}]{woods2014mappings}%
  \BibitemOpen
  \bibfield  {author} {\bibinfo {author} {\bibfnamefont {M.~P.}\ \bibnamefont
  {Woods}}, \bibinfo {author} {\bibfnamefont {R.}~\bibnamefont {Groux}},
  \bibinfo {author} {\bibfnamefont {A.~W.}\ \bibnamefont {Chin}}, \bibinfo
  {author} {\bibfnamefont {S.~F.}\ \bibnamefont {Huelga}},\ and\ \bibinfo
  {author} {\bibfnamefont {M.~B.}\ \bibnamefont {Plenio}},\ }\bibfield  {title}
  {\bibinfo {title} {Mappings of open quantum systems onto chain
  representations and {Markovian} embeddings},\ }\href@noop {} {\bibfield
  {journal} {\bibinfo  {journal} {J. Math. Phys.}\ }\textbf {\bibinfo {volume}
  {55}},\ \bibinfo {pages} {032101} (\bibinfo {year} {2014})}\BibitemShut
  {NoStop}%
\bibitem [{\citenamefont {Huh}\ \emph {et~al.}(2014)\citenamefont {Huh},
  \citenamefont {Mostame}, \citenamefont {Fujita}, \citenamefont {Yung},\ and\
  \citenamefont {Aspuru-Guzik}}]{huh2014linear}%
  \BibitemOpen
  \bibfield  {author} {\bibinfo {author} {\bibfnamefont {J.}~\bibnamefont
  {Huh}}, \bibinfo {author} {\bibfnamefont {S.}~\bibnamefont {Mostame}},
  \bibinfo {author} {\bibfnamefont {T.}~\bibnamefont {Fujita}}, \bibinfo
  {author} {\bibfnamefont {M.-H.}\ \bibnamefont {Yung}},\ and\ \bibinfo
  {author} {\bibfnamefont {A.}~\bibnamefont {Aspuru-Guzik}},\ }\bibfield
  {title} {\bibinfo {title} {Linear-algebraic bath transformation for
  simulating complex open quantum systems},\ }\href@noop {} {\bibfield
  {journal} {\bibinfo  {journal} {New J. Phys.}\ }\textbf {\bibinfo {volume}
  {16}},\ \bibinfo {pages} {123008} (\bibinfo {year} {2014})}\BibitemShut
  {NoStop}%
\bibitem [{\citenamefont {N{\"u}{\ss}eler}\ \emph {et~al.}(2022)\citenamefont
  {N{\"u}{\ss}eler}, \citenamefont {Tamascelli}, \citenamefont {Smirne},
  \citenamefont {Lim}, \citenamefont {Huelga},\ and\ \citenamefont
  {Plenio}}]{nusseler2022fingerprint}%
  \BibitemOpen
  \bibfield  {author} {\bibinfo {author} {\bibfnamefont {A.}~\bibnamefont
  {N{\"u}{\ss}eler}}, \bibinfo {author} {\bibfnamefont {D.}~\bibnamefont
  {Tamascelli}}, \bibinfo {author} {\bibfnamefont {A.}~\bibnamefont {Smirne}},
  \bibinfo {author} {\bibfnamefont {J.}~\bibnamefont {Lim}}, \bibinfo {author}
  {\bibfnamefont {S.~F.}\ \bibnamefont {Huelga}},\ and\ \bibinfo {author}
  {\bibfnamefont {M.~B.}\ \bibnamefont {Plenio}},\ }\bibfield  {title}
  {\bibinfo {title} {Fingerprint and universal {Markovian} closure of
  structured bosonic environments},\ }\href@noop {} {\bibfield  {journal}
  {\bibinfo  {journal} {Phys. Rev. Lett.}\ }\textbf {\bibinfo {volume} {129}},\
  \bibinfo {pages} {140604} (\bibinfo {year} {2022})}\BibitemShut {NoStop}%
\bibitem [{\citenamefont {Prior}\ \emph {et~al.}(2010)\citenamefont {Prior},
  \citenamefont {Chin}, \citenamefont {Huelga},\ and\ \citenamefont
  {Plenio}}]{prior2010efficient}%
  \BibitemOpen
  \bibfield  {author} {\bibinfo {author} {\bibfnamefont {J.}~\bibnamefont
  {Prior}}, \bibinfo {author} {\bibfnamefont {A.~W.}\ \bibnamefont {Chin}},
  \bibinfo {author} {\bibfnamefont {S.~F.}\ \bibnamefont {Huelga}},\ and\
  \bibinfo {author} {\bibfnamefont {M.~B.}\ \bibnamefont {Plenio}},\ }\bibfield
   {title} {\bibinfo {title} {Efficient simulation of strong system-environment
  interactions},\ }\href@noop {} {\bibfield  {journal} {\bibinfo  {journal}
  {Phys. Rev. Lett.}\ }\textbf {\bibinfo {volume} {105}},\ \bibinfo {pages}
  {050404} (\bibinfo {year} {2010})}\BibitemShut {NoStop}%
\bibitem [{\citenamefont {de~Vega}\ and\ \citenamefont
  {Ba{\~n}uls}(2015)}]{de2015thermofield}%
  \BibitemOpen
  \bibfield  {author} {\bibinfo {author} {\bibfnamefont {I.}~\bibnamefont
  {de~Vega}}\ and\ \bibinfo {author} {\bibfnamefont {M.-C.}\ \bibnamefont
  {Ba{\~n}uls}},\ }\bibfield  {title} {\bibinfo {title} {Thermofield-based
  chain-mapping approach for open quantum systems},\ }\href@noop {} {\bibfield
  {journal} {\bibinfo  {journal} {Phys. Rev. A}\ }\textbf {\bibinfo {volume}
  {92}},\ \bibinfo {pages} {052116} (\bibinfo {year} {2015})}\BibitemShut
  {NoStop}%
\bibitem [{\citenamefont {Tamascelli}\ \emph {et~al.}(2019)\citenamefont
  {Tamascelli}, \citenamefont {Smirne}, \citenamefont {Lim}, \citenamefont
  {Huelga},\ and\ \citenamefont {Plenio}}]{tamascelli2019efficient}%
  \BibitemOpen
  \bibfield  {author} {\bibinfo {author} {\bibfnamefont {D.}~\bibnamefont
  {Tamascelli}}, \bibinfo {author} {\bibfnamefont {A.}~\bibnamefont {Smirne}},
  \bibinfo {author} {\bibfnamefont {J.}~\bibnamefont {Lim}}, \bibinfo {author}
  {\bibfnamefont {S.~F.}\ \bibnamefont {Huelga}},\ and\ \bibinfo {author}
  {\bibfnamefont {M.~B.}\ \bibnamefont {Plenio}},\ }\bibfield  {title}
  {\bibinfo {title} {Efficient simulation of finite-temperature open quantum
  systems},\ }\href@noop {} {\bibfield  {journal} {\bibinfo  {journal} {Phys.
  Rev. Lett.}\ }\textbf {\bibinfo {volume} {123}},\ \bibinfo {pages} {090402}
  (\bibinfo {year} {2019})}\BibitemShut {NoStop}%
\bibitem [{\citenamefont {Halliwell}\ and\ \citenamefont
  {Yu}(1996)}]{halliwell1996alternative}%
  \BibitemOpen
  \bibfield  {author} {\bibinfo {author} {\bibfnamefont {J.~J.}\ \bibnamefont
  {Halliwell}}\ and\ \bibinfo {author} {\bibfnamefont {T.}~\bibnamefont {Yu}},\
  }\bibfield  {title} {\bibinfo {title} {Alternative derivation of the
  {Hu-Paz-Zhang} master equation of quantum {Brownian} motion},\ }\href@noop {}
  {\bibfield  {journal} {\bibinfo  {journal} {Phys. Rev. D}\ }\textbf {\bibinfo
  {volume} {53}},\ \bibinfo {pages} {2012} (\bibinfo {year}
  {1996})}\BibitemShut {NoStop}%
\bibitem [{\citenamefont {Chou}\ \emph
  {et~al.}(2008{\natexlab{a}})\citenamefont {Chou}, \citenamefont {Hu},\ and\
  \citenamefont {Yu}}]{chou2008quantum}%
  \BibitemOpen
  \bibfield  {author} {\bibinfo {author} {\bibfnamefont {C.-H.}\ \bibnamefont
  {Chou}}, \bibinfo {author} {\bibfnamefont {B.~L.}\ \bibnamefont {Hu}},\ and\
  \bibinfo {author} {\bibfnamefont {T.}~\bibnamefont {Yu}},\ }\bibfield
  {title} {\bibinfo {title} {Quantum {Brownian} motion of a macroscopic object
  in a general environment},\ }\href@noop {} {\bibfield  {journal} {\bibinfo
  {journal} {Physica A}\ }\textbf {\bibinfo {volume} {387}},\ \bibinfo {pages}
  {432} (\bibinfo {year} {2008}{\natexlab{a}})}\BibitemShut {NoStop}%
\bibitem [{\citenamefont {Chou}\ \emph
  {et~al.}(2008{\natexlab{b}})\citenamefont {Chou}, \citenamefont {Yu},\ and\
  \citenamefont {Hu}}]{chou2008exact}%
  \BibitemOpen
  \bibfield  {author} {\bibinfo {author} {\bibfnamefont {C.~H.}\ \bibnamefont
  {Chou}}, \bibinfo {author} {\bibfnamefont {T.}~\bibnamefont {Yu}},\ and\
  \bibinfo {author} {\bibfnamefont {B.~L.}\ \bibnamefont {Hu}},\ }\bibfield
  {title} {\bibinfo {title} {Exact master equation and quantum decoherence of
  two coupled harmonic oscillators in a general environment},\ }\href@noop {}
  {\bibfield  {journal} {\bibinfo  {journal} {Phys. Rev. E}\ }\textbf {\bibinfo
  {volume} {77}},\ \bibinfo {pages} {011112} (\bibinfo {year}
  {2008}{\natexlab{b}})}\BibitemShut {NoStop}%
\bibitem [{\citenamefont {Martinez}\ and\ \citenamefont
  {Paz}(2013)}]{martinez2013dynamics}%
  \BibitemOpen
  \bibfield  {author} {\bibinfo {author} {\bibfnamefont {E.~A.}\ \bibnamefont
  {Martinez}}\ and\ \bibinfo {author} {\bibfnamefont {J.~P.}\ \bibnamefont
  {Paz}},\ }\bibfield  {title} {\bibinfo {title} {Dynamics and thermodynamics
  of linear quantum open systems},\ }\href@noop {} {\bibfield  {journal}
  {\bibinfo  {journal} {Phys. Rev. Lett.}\ }\textbf {\bibinfo {volume} {110}},\
  \bibinfo {pages} {130406} (\bibinfo {year} {2013})}\BibitemShut {NoStop}%
\bibitem [{\citenamefont {Jauho}\ \emph {et~al.}(1994)\citenamefont {Jauho},
  \citenamefont {Wingreen},\ and\ \citenamefont {Meir}}]{jauho1994time}%
  \BibitemOpen
  \bibfield  {author} {\bibinfo {author} {\bibfnamefont {A.-P.}\ \bibnamefont
  {Jauho}}, \bibinfo {author} {\bibfnamefont {N.~S.}\ \bibnamefont
  {Wingreen}},\ and\ \bibinfo {author} {\bibfnamefont {Y.}~\bibnamefont
  {Meir}},\ }\bibfield  {title} {\bibinfo {title} {Time-dependent transport in
  interacting and noninteracting resonant-tunneling systems},\ }\href@noop {}
  {\bibfield  {journal} {\bibinfo  {journal} {Phys. Rev. B}\ }\textbf {\bibinfo
  {volume} {50}},\ \bibinfo {pages} {5528} (\bibinfo {year}
  {1994})}\BibitemShut {NoStop}%
\bibitem [{\citenamefont {Dhar}\ and\ \citenamefont
  {Roy}(2006)}]{dhar2006heat}%
  \BibitemOpen
  \bibfield  {author} {\bibinfo {author} {\bibfnamefont {A.}~\bibnamefont
  {Dhar}}\ and\ \bibinfo {author} {\bibfnamefont {D.}~\bibnamefont {Roy}},\
  }\bibfield  {title} {\bibinfo {title} {Heat transport in harmonic lattices},\
  }\href@noop {} {\bibfield  {journal} {\bibinfo  {journal} {J. Stat. Phys.}\
  }\textbf {\bibinfo {volume} {125}},\ \bibinfo {pages} {801} (\bibinfo {year}
  {2006})}\BibitemShut {NoStop}%
\bibitem [{\citenamefont {Jin}\ \emph {et~al.}(2010)\citenamefont {Jin},
  \citenamefont {Tu}, \citenamefont {Zhang},\ and\ \citenamefont
  {Yan}}]{jin2010non}%
  \BibitemOpen
  \bibfield  {author} {\bibinfo {author} {\bibfnamefont {J.}~\bibnamefont
  {Jin}}, \bibinfo {author} {\bibfnamefont {M.~W.-Y.}\ \bibnamefont {Tu}},
  \bibinfo {author} {\bibfnamefont {W.-M.}\ \bibnamefont {Zhang}},\ and\
  \bibinfo {author} {\bibfnamefont {Y.}~\bibnamefont {Yan}},\ }\bibfield
  {title} {\bibinfo {title} {Non-equilibrium quantum theory for nanodevices
  based on the {Feynman--Vernon} influence functional},\ }\href@noop {}
  {\bibfield  {journal} {\bibinfo  {journal} {New J. Phys.}\ }\textbf {\bibinfo
  {volume} {12}},\ \bibinfo {pages} {083013} (\bibinfo {year}
  {2010})}\BibitemShut {NoStop}%
\bibitem [{\citenamefont {Alghadeer}\ \emph {et~al.}(2026)\citenamefont
  {Alghadeer}, \citenamefont {Pettersson~Fors}, \citenamefont {Cao},
  \citenamefont {Fasciati}, \citenamefont {Ishizaka}, \citenamefont {Kockum},
  \citenamefont {Leek},\ and\ \citenamefont {Bakr}}]{alghadeer2026crosstalk}%
  \BibitemOpen
  \bibfield  {author} {\bibinfo {author} {\bibfnamefont {M.}~\bibnamefont
  {Alghadeer}}, \bibinfo {author} {\bibfnamefont {S.}~\bibnamefont
  {Pettersson~Fors}}, \bibinfo {author} {\bibfnamefont {S.}~\bibnamefont
  {Cao}}, \bibinfo {author} {\bibfnamefont {S.~D.}\ \bibnamefont {Fasciati}},
  \bibinfo {author} {\bibfnamefont {H.}~\bibnamefont {Ishizaka}}, \bibinfo
  {author} {\bibfnamefont {A.~F.}\ \bibnamefont {Kockum}}, \bibinfo {author}
  {\bibfnamefont {P.}~\bibnamefont {Leek}},\ and\ \bibinfo {author}
  {\bibfnamefont {M.}~\bibnamefont {Bakr}},\ }\bibfield  {title} {\bibinfo
  {title} {Crosstalk dispersion and spatial scaling in superconducting qubit
  arrays},\ }\href@noop {} {\bibfield  {journal} {\bibinfo  {journal} {Phys.
  Rev. Research}\ }\textbf {\bibinfo {volume} {8}},\ \bibinfo {pages} {033193}
  (\bibinfo {year} {2026})}\BibitemShut {NoStop}%
\bibitem [{\citenamefont {Hughes}\ \emph {et~al.}(2009)\citenamefont {Hughes},
  \citenamefont {Christ},\ and\ \citenamefont
  {Burghardt}}]{hughes2009effective}%
  \BibitemOpen
  \bibfield  {author} {\bibinfo {author} {\bibfnamefont {K.~H.}\ \bibnamefont
  {Hughes}}, \bibinfo {author} {\bibfnamefont {C.~D.}\ \bibnamefont {Christ}},\
  and\ \bibinfo {author} {\bibfnamefont {I.}~\bibnamefont {Burghardt}},\
  }\bibfield  {title} {\bibinfo {title} {Effective-mode representation of
  non-{Markovian} dynamics: A hierarchical approximation of the spectral
  density. {I}. {Application} to single surface dynamics},\ }\href@noop {}
  {\bibfield  {journal} {\bibinfo  {journal} {J. Chem. Phys.}\ }\textbf
  {\bibinfo {volume} {131}},\ \bibinfo {pages} {024109} (\bibinfo {year}
  {2009})}\BibitemShut {NoStop}%
\bibitem [{\citenamefont {Martinazzo}\ \emph
  {et~al.}(2011{\natexlab{b}})\citenamefont {Martinazzo}, \citenamefont
  {Hughes},\ and\ \citenamefont {Burghardt}}]{martinazzo2011unraveling}%
  \BibitemOpen
  \bibfield  {author} {\bibinfo {author} {\bibfnamefont {R.}~\bibnamefont
  {Martinazzo}}, \bibinfo {author} {\bibfnamefont {K.~H.}\ \bibnamefont
  {Hughes}},\ and\ \bibinfo {author} {\bibfnamefont {I.}~\bibnamefont
  {Burghardt}},\ }\bibfield  {title} {\bibinfo {title} {Unraveling a {Brownian}
  particle’s memory with effective mode chains},\ }\href@noop {} {\bibfield
  {journal} {\bibinfo  {journal} {Phys. Rev. E}\ }\textbf {\bibinfo {volume}
  {84}},\ \bibinfo {pages} {030102} (\bibinfo {year}
  {2011}{\natexlab{b}})}\BibitemShut {NoStop}%
\bibitem [{\citenamefont {Bakr}\ and\ \citenamefont
  {Wopalenski}(2026)}]{bakr2026exact}%
  \BibitemOpen
  \bibfield  {author} {\bibinfo {author} {\bibfnamefont {M.}~\bibnamefont
  {Bakr}}\ and\ \bibinfo {author} {\bibfnamefont {R.}~\bibnamefont
  {Wopalenski}},\ }\bibfield  {title} {\bibinfo {title} {Exact multimode
  quantization of superconducting circuits via boundary admittance and
  continued fractions},\ }\href@noop {} {\bibfield  {journal} {\bibinfo
  {journal} {arXiv preprint arXiv:2601.04407}\ } (\bibinfo {year}
  {2026})}\BibitemShut {NoStop}%
\bibitem [{\citenamefont {Caroli}\ \emph {et~al.}(1971)\citenamefont {Caroli},
  \citenamefont {Combescot}, \citenamefont {Nozi{\`e}res},\ and\ \citenamefont
  {Saint-James}}]{caroli1971direct}%
  \BibitemOpen
  \bibfield  {author} {\bibinfo {author} {\bibfnamefont {C.}~\bibnamefont
  {Caroli}}, \bibinfo {author} {\bibfnamefont {R.}~\bibnamefont {Combescot}},
  \bibinfo {author} {\bibfnamefont {P.}~\bibnamefont {Nozi{\`e}res}},\ and\
  \bibinfo {author} {\bibfnamefont {D.}~\bibnamefont {Saint-James}},\
  }\bibfield  {title} {\bibinfo {title} {Direct calculation of the tunneling
  current},\ }\href {https://doi.org/10.1088/0022-3719/4/8/018} {\bibfield
  {journal} {\bibinfo  {journal} {Journal of Physics C: Solid State Physics}\
  }\textbf {\bibinfo {volume} {4}},\ \bibinfo {pages} {916} (\bibinfo {year}
  {1971})}\BibitemShut {NoStop}%
\bibitem [{\citenamefont {Lake}\ \emph {et~al.}(1997)\citenamefont {Lake},
  \citenamefont {Klimeck}, \citenamefont {Bowen},\ and\ \citenamefont
  {Jovanovic}}]{lake1997single}%
  \BibitemOpen
  \bibfield  {author} {\bibinfo {author} {\bibfnamefont {R.}~\bibnamefont
  {Lake}}, \bibinfo {author} {\bibfnamefont {G.}~\bibnamefont {Klimeck}},
  \bibinfo {author} {\bibfnamefont {R.~C.}\ \bibnamefont {Bowen}},\ and\
  \bibinfo {author} {\bibfnamefont {D.}~\bibnamefont {Jovanovic}},\ }\bibfield
  {title} {\bibinfo {title} {Single and multiband modeling of quantum electron
  transport through layered semiconductor devices},\ }\href@noop {} {\bibfield
  {journal} {\bibinfo  {journal} {Journal of Applied Physics}\ }\textbf
  {\bibinfo {volume} {81}},\ \bibinfo {pages} {7845} (\bibinfo {year}
  {1997})}\BibitemShut {NoStop}%
\bibitem [{\citenamefont {Luisier}\ \emph {et~al.}(2006)\citenamefont
  {Luisier}, \citenamefont {Schenk}, \citenamefont {Fichtner},\ and\
  \citenamefont {Klimeck}}]{luisier2006atomistic}%
  \BibitemOpen
  \bibfield  {author} {\bibinfo {author} {\bibfnamefont {M.}~\bibnamefont
  {Luisier}}, \bibinfo {author} {\bibfnamefont {A.}~\bibnamefont {Schenk}},
  \bibinfo {author} {\bibfnamefont {W.}~\bibnamefont {Fichtner}},\ and\
  \bibinfo {author} {\bibfnamefont {G.}~\bibnamefont {Klimeck}},\ }\bibfield
  {title} {\bibinfo {title} {Atomistic simulation of nanowires in the sp 3 d 5
  s* tight-binding formalism: From boundary conditions to strain
  calculations},\ }\href@noop {} {\bibfield  {journal} {\bibinfo  {journal}
  {Physical Review B—Condensed Matter and Materials Physics}\ }\textbf
  {\bibinfo {volume} {74}},\ \bibinfo {pages} {205323} (\bibinfo {year}
  {2006})}\BibitemShut {NoStop}%
\bibitem [{\citenamefont {Luisier}\ and\ \citenamefont
  {Klimeck}(2009)}]{luisier2009atomistic}%
  \BibitemOpen
  \bibfield  {author} {\bibinfo {author} {\bibfnamefont {M.}~\bibnamefont
  {Luisier}}\ and\ \bibinfo {author} {\bibfnamefont {G.}~\bibnamefont
  {Klimeck}},\ }\bibfield  {title} {\bibinfo {title} {Atomistic full-band
  simulations of silicon nanowire transistors: Effects of electron-phonon
  scattering},\ }\href@noop {} {\bibfield  {journal} {\bibinfo  {journal}
  {Physical Review B—Condensed Matter and Materials Physics}\ }\textbf
  {\bibinfo {volume} {80}},\ \bibinfo {pages} {155430} (\bibinfo {year}
  {2009})}\BibitemShut {NoStop}%
\bibitem [{\citenamefont {Haydock}\ \emph {et~al.}(1972)\citenamefont
  {Haydock}, \citenamefont {Heine},\ and\ \citenamefont
  {Kelly}}]{haydock1972electronic}%
  \BibitemOpen
  \bibfield  {author} {\bibinfo {author} {\bibfnamefont {R.}~\bibnamefont
  {Haydock}}, \bibinfo {author} {\bibfnamefont {V.}~\bibnamefont {Heine}},\
  and\ \bibinfo {author} {\bibfnamefont {M.}~\bibnamefont {Kelly}},\ }\bibfield
   {title} {\bibinfo {title} {Electronic structure based on the local atomic
  environment for tight-binding bands},\ }\href@noop {} {\bibfield  {journal}
  {\bibinfo  {journal} {Journal of Physics C: Solid State Physics}\ }\textbf
  {\bibinfo {volume} {5}},\ \bibinfo {pages} {2845} (\bibinfo {year}
  {1972})}\BibitemShut {NoStop}%
\bibitem [{\citenamefont {Haydock}(1980)}]{haydock1980recursive}%
  \BibitemOpen
  \bibfield  {author} {\bibinfo {author} {\bibfnamefont {R.}~\bibnamefont
  {Haydock}},\ }\bibfield  {title} {\bibinfo {title} {The recursive solution of
  the {Schr{\"o}dinger} equation},\ }in\ \href@noop {} {\emph {\bibinfo
  {booktitle} {Solid State Physics}}},\ Vol.~\bibinfo {volume} {35},\ \bibinfo
  {editor} {edited by\ \bibinfo {editor} {\bibfnamefont {H.}~\bibnamefont
  {Ehrenreich}}, \bibinfo {editor} {\bibfnamefont {F.}~\bibnamefont {Seitz}},\
  and\ \bibinfo {editor} {\bibfnamefont {D.}~\bibnamefont {Turnbull}}}\
  (\bibinfo  {publisher} {Academic Press},\ \bibinfo {address} {New York},\
  \bibinfo {year} {1980})\ pp.\ \bibinfo {pages} {215--294}\BibitemShut
  {NoStop}%
\bibitem [{\citenamefont {Beer}\ and\ \citenamefont
  {Pettifor}(1984)}]{beer1984recursion}%
  \BibitemOpen
  \bibfield  {author} {\bibinfo {author} {\bibfnamefont {N.}~\bibnamefont
  {Beer}}\ and\ \bibinfo {author} {\bibfnamefont {D.~G.}\ \bibnamefont
  {Pettifor}},\ }\bibfield  {title} {\bibinfo {title} {The recursion method and
  the estimation of local densities of states},\ }in\ \href@noop {} {\emph
  {\bibinfo {booktitle} {The Electronic Structure of Complex Systems}}},\
  \bibinfo {series} {NATO ASI Series}, Vol.\ \bibinfo {volume} {113},\ \bibinfo
  {editor} {edited by\ \bibinfo {editor} {\bibfnamefont {P.}~\bibnamefont
  {Phariseau}}\ and\ \bibinfo {editor} {\bibfnamefont {W.~M.}\ \bibnamefont
  {Temmerman}}}\ (\bibinfo  {publisher} {Plenum Press},\ \bibinfo {address}
  {New York},\ \bibinfo {year} {1984})\ pp.\ \bibinfo {pages}
  {769--777}\BibitemShut {NoStop}%
\bibitem [{\citenamefont {Kuijlaars}\ \emph {et~al.}(2004)\citenamefont
  {Kuijlaars}, \citenamefont {McLaughlin}, \citenamefont {Van~Assche},\ and\
  \citenamefont {Vanlessen}}]{kuijlaars2004riemann}%
  \BibitemOpen
  \bibfield  {author} {\bibinfo {author} {\bibfnamefont {A.~B.}\ \bibnamefont
  {Kuijlaars}}, \bibinfo {author} {\bibfnamefont {K.-R.}\ \bibnamefont
  {McLaughlin}}, \bibinfo {author} {\bibfnamefont {W.}~\bibnamefont
  {Van~Assche}},\ and\ \bibinfo {author} {\bibfnamefont {M.}~\bibnamefont
  {Vanlessen}},\ }\bibfield  {title} {\bibinfo {title} {The
  {R}iemann--{H}ilbert approach to strong asymptotics for orthogonal
  polynomials on $[-1,1]$},\ }\href@noop {} {\bibfield  {journal} {\bibinfo
  {journal} {Advances in Mathematics}\ }\textbf {\bibinfo {volume} {188}},\
  \bibinfo {pages} {337} (\bibinfo {year} {2004})}\BibitemShut {NoStop}%
\bibitem [{\citenamefont {Leggett}\ \emph {et~al.}(1987)\citenamefont
  {Leggett}, \citenamefont {Chakravarty}, \citenamefont {Dorsey}, \citenamefont
  {Fisher}, \citenamefont {Garg},\ and\ \citenamefont
  {Zwerger}}]{leggett1987dynamics}%
  \BibitemOpen
  \bibfield  {author} {\bibinfo {author} {\bibfnamefont {A.~J.}\ \bibnamefont
  {Leggett}}, \bibinfo {author} {\bibfnamefont {S.}~\bibnamefont
  {Chakravarty}}, \bibinfo {author} {\bibfnamefont {A.~T.}\ \bibnamefont
  {Dorsey}}, \bibinfo {author} {\bibfnamefont {M.~P.}\ \bibnamefont {Fisher}},
  \bibinfo {author} {\bibfnamefont {A.}~\bibnamefont {Garg}},\ and\ \bibinfo
  {author} {\bibfnamefont {W.}~\bibnamefont {Zwerger}},\ }\bibfield  {title}
  {\bibinfo {title} {Dynamics of the dissipative two-state system},\
  }\href@noop {} {\bibfield  {journal} {\bibinfo  {journal} {Reviews of Modern
  Physics}\ }\textbf {\bibinfo {volume} {59}},\ \bibinfo {pages} {1} (\bibinfo
  {year} {1987})}\BibitemShut {NoStop}%
\bibitem [{\citenamefont {Weiss}(2012)}]{weiss2012quantum}%
  \BibitemOpen
  \bibfield  {author} {\bibinfo {author} {\bibfnamefont {U.}~\bibnamefont
  {Weiss}},\ }\href@noop {} {\emph {\bibinfo {title} {Quantum dissipative
  systems}}}\ (\bibinfo  {publisher} {World Scientific},\ \bibinfo {year}
  {2012})\BibitemShut {NoStop}%
\bibitem [{\citenamefont {Garraway}(1997)}]{garraway1997nonperturbative}%
  \BibitemOpen
  \bibfield  {author} {\bibinfo {author} {\bibfnamefont {B.}~\bibnamefont
  {Garraway}},\ }\bibfield  {title} {\bibinfo {title} {Nonperturbative decay of
  an atomic system in a cavity},\ }\href@noop {} {\bibfield  {journal}
  {\bibinfo  {journal} {Physical Review A}\ }\textbf {\bibinfo {volume} {55}},\
  \bibinfo {pages} {2290} (\bibinfo {year} {1997})}\BibitemShut {NoStop}%
\bibitem [{\citenamefont {Lambropoulos}\ \emph {et~al.}(2000)\citenamefont
  {Lambropoulos}, \citenamefont {Nikolopoulos}, \citenamefont {Nielsen},\ and\
  \citenamefont {Bay}}]{lambropoulos2000fundamental}%
  \BibitemOpen
  \bibfield  {author} {\bibinfo {author} {\bibfnamefont {P.}~\bibnamefont
  {Lambropoulos}}, \bibinfo {author} {\bibfnamefont {G.~M.}\ \bibnamefont
  {Nikolopoulos}}, \bibinfo {author} {\bibfnamefont {T.~R.}\ \bibnamefont
  {Nielsen}},\ and\ \bibinfo {author} {\bibfnamefont {S.}~\bibnamefont {Bay}},\
  }\bibfield  {title} {\bibinfo {title} {Fundamental quantum optics in
  structured reservoirs},\ }\href@noop {} {\bibfield  {journal} {\bibinfo
  {journal} {Rep. Prog. Phys.}\ }\textbf {\bibinfo {volume} {63}},\ \bibinfo
  {pages} {455} (\bibinfo {year} {2000})}\BibitemShut {NoStop}%
\bibitem [{\citenamefont {Cohen-Tannoudji}\ \emph {et~al.}(2024)\citenamefont
  {Cohen-Tannoudji}, \citenamefont {Dupont-Roc},\ and\ \citenamefont
  {Grynberg}}]{cohen2024atom}%
  \BibitemOpen
  \bibfield  {author} {\bibinfo {author} {\bibfnamefont {C.}~\bibnamefont
  {Cohen-Tannoudji}}, \bibinfo {author} {\bibfnamefont {J.}~\bibnamefont
  {Dupont-Roc}},\ and\ \bibinfo {author} {\bibfnamefont {G.}~\bibnamefont
  {Grynberg}},\ }\href@noop {} {\emph {\bibinfo {title} {Atom-photon
  interactions: basic processes and applications}}}\ (\bibinfo  {publisher}
  {John Wiley \& Sons},\ \bibinfo {year} {2024})\BibitemShut {NoStop}%
\bibitem [{\citenamefont {Wigner}(1955)}]{wigner1955characteristic}%
  \BibitemOpen
  \bibfield  {author} {\bibinfo {author} {\bibfnamefont {E.~P.}\ \bibnamefont
  {Wigner}},\ }\bibfield  {title} {\bibinfo {title} {Characteristic vectors of
  bordered matrices with infinite dimensions},\ }\href
  {https://doi.org/10.2307/1970079} {\bibfield  {journal} {\bibinfo  {journal}
  {Ann. Math.}\ }\textbf {\bibinfo {volume} {62}},\ \bibinfo {pages} {548}
  (\bibinfo {year} {1955})}\BibitemShut {NoStop}%
\bibitem [{\citenamefont {Wigner}(1958)}]{wigner1958distribution}%
  \BibitemOpen
  \bibfield  {author} {\bibinfo {author} {\bibfnamefont {E.~P.}\ \bibnamefont
  {Wigner}},\ }\bibfield  {title} {\bibinfo {title} {On the distribution of the
  roots of certain symmetric matrices},\ }\href@noop {} {\bibfield  {journal}
  {\bibinfo  {journal} {Annals of Mathematics}\ }\textbf {\bibinfo {volume}
  {67}},\ \bibinfo {pages} {325} (\bibinfo {year} {1958})}\BibitemShut
  {NoStop}%
\bibitem [{\citenamefont {Pastur}(1972)}]{pastur1972spectrum}%
  \BibitemOpen
  \bibfield  {author} {\bibinfo {author} {\bibfnamefont {L.~A.}\ \bibnamefont
  {Pastur}},\ }\bibfield  {title} {\bibinfo {title} {On the spectrum of random
  matrices},\ }\href {https://doi.org/10.1007/BF01035768} {\bibfield  {journal}
  {\bibinfo  {journal} {Theor. Math. Phys.}\ }\textbf {\bibinfo {volume}
  {10}},\ \bibinfo {pages} {67} (\bibinfo {year} {1972})}\BibitemShut {NoStop}%
\bibitem [{\citenamefont {M{\"u}ller}\ and\ \citenamefont
  {Strunz}(2026)}]{muller2026one}%
  \BibitemOpen
  \bibfield  {author} {\bibinfo {author} {\bibfnamefont {K.}~\bibnamefont
  {M{\"u}ller}}\ and\ \bibinfo {author} {\bibfnamefont {W.~T.}\ \bibnamefont
  {Strunz}},\ }\bibfield  {title} {\bibinfo {title} {One-to-one correspondence
  between hierarchical equations of motion and pseudomodes for open quantum
  system dynamics},\ }\href@noop {} {\bibfield  {journal} {\bibinfo  {journal}
  {arXiv preprint arXiv:2604.06466}\ } (\bibinfo {year} {2026})}\BibitemShut
  {NoStop}%
\bibitem [{\citenamefont {Garnett}(2007)}]{garnett2007bounded}%
  \BibitemOpen
  \bibfield  {author} {\bibinfo {author} {\bibfnamefont {J.~B.}\ \bibnamefont
  {Garnett}},\ }\href@noop {} {\emph {\bibinfo {title} {Bounded Analytic
  Functions}}},\ \bibinfo {edition} {revised first}\ ed.,\ \bibinfo {series}
  {Graduate Texts in Mathematics}, Vol.\ \bibinfo {volume} {236}\ (\bibinfo
  {publisher} {Springer},\ \bibinfo {address} {New York},\ \bibinfo {year}
  {2007})\BibitemShut {NoStop}%
\bibitem [{\citenamefont {Tamascelli}\ \emph {et~al.}(2018)\citenamefont
  {Tamascelli}, \citenamefont {Smirne}, \citenamefont {Huelga},\ and\
  \citenamefont {Plenio}}]{tamascelli2018nonperturbative}%
  \BibitemOpen
  \bibfield  {author} {\bibinfo {author} {\bibfnamefont {D.}~\bibnamefont
  {Tamascelli}}, \bibinfo {author} {\bibfnamefont {A.}~\bibnamefont {Smirne}},
  \bibinfo {author} {\bibfnamefont {S.~F.}\ \bibnamefont {Huelga}},\ and\
  \bibinfo {author} {\bibfnamefont {M.~B.}\ \bibnamefont {Plenio}},\ }\bibfield
   {title} {\bibinfo {title} {Nonperturbative treatment of non-{M}arkovian
  dynamics of open quantum systems},\ }\href@noop {} {\bibfield  {journal}
  {\bibinfo  {journal} {Physical Review Letters}\ }\textbf {\bibinfo {volume}
  {120}},\ \bibinfo {pages} {030402} (\bibinfo {year} {2018})}\BibitemShut
  {NoStop}%
\bibitem [{\citenamefont {Papaefstathiou}\ \emph {et~al.}(2025)\citenamefont
  {Papaefstathiou}, \citenamefont {Malz}, \citenamefont {Cirac},\ and\
  \citenamefont {Ba{\~n}uls}}]{papaefstathiou2025efficient}%
  \BibitemOpen
  \bibfield  {author} {\bibinfo {author} {\bibfnamefont {I.}~\bibnamefont
  {Papaefstathiou}}, \bibinfo {author} {\bibfnamefont {D.}~\bibnamefont
  {Malz}}, \bibinfo {author} {\bibfnamefont {J.~I.}\ \bibnamefont {Cirac}},\
  and\ \bibinfo {author} {\bibfnamefont {M.~C.}\ \bibnamefont {Ba{\~n}uls}},\
  }\bibfield  {title} {\bibinfo {title} {Efficient tensor-network simulation of
  multiemitter non-{M}arkovian systems},\ }\href@noop {} {\bibfield  {journal}
  {\bibinfo  {journal} {Physical Review A}\ }\textbf {\bibinfo {volume}
  {112}},\ \bibinfo {pages} {013721} (\bibinfo {year} {2025})}\BibitemShut
  {NoStop}%
\bibitem [{\citenamefont {Sundaresan}\ \emph {et~al.}(2019)\citenamefont
  {Sundaresan}, \citenamefont {Lundgren}, \citenamefont {Zhu}, \citenamefont
  {Gorshkov},\ and\ \citenamefont {Houck}}]{sundaresan2019interacting}%
  \BibitemOpen
  \bibfield  {author} {\bibinfo {author} {\bibfnamefont {N.~M.}\ \bibnamefont
  {Sundaresan}}, \bibinfo {author} {\bibfnamefont {R.}~\bibnamefont
  {Lundgren}}, \bibinfo {author} {\bibfnamefont {G.}~\bibnamefont {Zhu}},
  \bibinfo {author} {\bibfnamefont {A.~V.}\ \bibnamefont {Gorshkov}},\ and\
  \bibinfo {author} {\bibfnamefont {A.~A.}\ \bibnamefont {Houck}},\ }\bibfield
  {title} {\bibinfo {title} {Interacting qubit-photon bound states with
  superconducting circuits},\ }\href@noop {} {\bibfield  {journal} {\bibinfo
  {journal} {Physical Review X}\ }\textbf {\bibinfo {volume} {9}},\ \bibinfo
  {pages} {011021} (\bibinfo {year} {2019})}\BibitemShut {NoStop}%
\bibitem [{\citenamefont {Viswanath}\ and\ \citenamefont
  {M{\"u}ller}(1994)}]{viswanath1994recursion}%
  \BibitemOpen
  \bibfield  {author} {\bibinfo {author} {\bibfnamefont {V.}~\bibnamefont
  {Viswanath}}\ and\ \bibinfo {author} {\bibfnamefont {G.}~\bibnamefont
  {M{\"u}ller}},\ }\href@noop {} {\emph {\bibinfo {title} {The recursion
  method: application to many-body dynamics}}}\ (\bibinfo  {publisher}
  {Springer},\ \bibinfo {year} {1994})\BibitemShut {NoStop}%
\bibitem [{\citenamefont {Gautschi}(2004)}]{gautschi2004orthogonal}%
  \BibitemOpen
  \bibfield  {author} {\bibinfo {author} {\bibfnamefont {W.}~\bibnamefont
  {Gautschi}},\ }\href@noop {} {\emph {\bibinfo {title} {Orthogonal
  polynomials: computation and approximation}}}\ (\bibinfo  {publisher} {OUP
  Oxford},\ \bibinfo {year} {2004})\BibitemShut {NoStop}%
\end{thebibliography}%

\end{document}